\documentclass[12pt]{article}
\usepackage{graphicx}
\usepackage{bm}
\usepackage{amsmath}
\usepackage{amssymb}
\usepackage{amsfonts}
\usepackage{mathtools}
\usepackage{hyperref}
\usepackage{braket}
\usepackage{anyfontsize}
\usepackage[normalem]{ulem}
\usepackage{wrapfig}
\usepackage{tikz}
\usepackage{cite}

\usepackage[left=1in,right=1in,top=1in,bottom=1in]{geometry}

\hypersetup{colorlinks=true,urlcolor=[rgb]{0,0,0.5},citecolor=[rgb]{0.5,0,0},linkcolor=[rgb]{0,0,0.4}}

\renewcommand{\title}[1]{\vbox{\center{\Large{#1}}}\vspace{5mm}}
\renewcommand{\author}[1]{\vbox{\center#1}\vspace{5mm}}
\newcommand{\address}[1]{\vbox{\center\em#1}}

\newcommand\emails[1]{\begingroup
	\renewcommand\thefootnote{}\footnote{#1}
	\addtocounter{footnote}{-1}\endgroup}

\def\Tr{{\rm Tr}}

\def\op{{\cal O}}
\def\C{\mathbb{C}}

\def\p{\partial}
\def\vev#1{\langle{#1}\rangle}
\def\iden{\mathbb{I}}
\def\1den{\hbox{$1\hskip -1.2pt\vrule depth 0pt height 1.53ex width 0.7pt
                  \vrule depth 0pt height 0.3pt width 0.12em$}}
\def\with{\quad {\rm with} \quad}
\def\where{\quad {\rm where} \quad}

\def\and{\quad {\rm and} \quad}

\def\nn{\nonumber\\}

\def\la{\leftarrow}
\def\ra{\rightarrow}
\def\ie{{\rm i.e.\ }}
\def\eg{{\rm e.g.\ }}

\def\CR{{\cal R}}

\def\Wg{{\cal W}\! g}

\begin{document}

\begin{titlepage}

\begin{center}
\vspace*{2cm}
\title{\bf Operator growth in random quantum circuits\\ with symmetry}
\author{Nicholas Hunter-Jones}
\address{
Institute for Quantum Information and Matter,\\ California Institute of Technology,\\ Pasadena, CA 91125, USA\\
\vspace*{4mm}
Perimeter Institute for Theoretical Physics,\\
Waterloo, ON N2L 2Y5, Canada}
\emails{ \hspace*{-8mm}
\href{mailto: nickrhj@perimeterinstitute.ca}{\tt nickrhj@perimeterinstitute.ca}}

\end{center}

\begin{abstract}
We study random quantum circuits with symmetry, where the local 2-site unitaries are drawn from a quotient or subgroup of the full unitary group $U(d)$. Random quantum circuits are minimal models of local quantum chaotic dynamics and can be used to study operator growth and the emergence of diffusive hydrodynamics. We derive the transition probabilities for the stochastic process governing the growth of operators in four classes of symmetric random circuits. We then compute the butterfly velocities and diffusion constants for a spreading operator by solving a simple random walk in each class of circuits.
\end{abstract}

\end{titlepage}

\tableofcontents

\section{Introduction}

Random quantum circuits have been studied extensively in the quantum information community, largely focused on understanding the convergence properties, for instance \cite{ELL05,ODP07,HarrowLow08,Brandao12,Zni08,BV10}. Specifically, random quantum circuits form approximate unitary $k$-designs in polynomial depth \cite{HarrowLow08,Brandao12}, rapidly scramble information \cite{Brown12} and achieve decoupling in polylogarithmic depth \cite{Brown15}. Recently, random quantum circuits have been used to study operator spreading under random unitary dynamics \cite{Nahum17,vonKey17} and with conservation laws \cite{Rakovszky17,Khemani17}, and give an emergent picture for the evolution of entanglement under unitary dynamics \cite{Nahum16,OpEnt18,RQCstatmech}. Operator growth is the statement that time evolution takes simple few-body operators to complicated combinations of non-local operators and has become understood as a symptom of chaotic dynamics in quantum many-body systems.\footnote{The growth of an operator also defines an effective causal structure in local systems \cite{LRbound,Hastings10} and is intimately related to scrambling  \cite{HaydenPreskill,SekinoSusskind,FastScrambling,LocalizedShocks,RobertsLR}. In this context, the study of ballistic operator growth has been explored in a wide variety of models\cite{Aleiner16,Gu16,PatelDiff,SwingleMPS,SYKopgrowth,ScrambGraphs18,SYKopbeta}.}

In this note we study operator growth in random quantum circuits with symmetry, where instead of Haar random 2-site unitaries, we construct the circuits using unitaries drawn from a quotient or subgroup of the unitary group. We consider four different symmetry classes of random circuits: orthogonal, symplectic, the circular orthogonal ensemble (COE), and circular symplectic ensemble (CSE). Each of the symmetric random circuits defines a different stochastic process of local updates on the evolving Pauli strings, which gives rise to operator growth under random circuit evolution. This constitutes four solvable models of local quantum chaotic dynamics with symmetry. 

The random quantum circuit models considered here are just a simple generalization of \cite{Nahum17,vonKey17}. Moreover, while our symmetric circuits builds in symmetry to the local random unitary, the random circuit as a whole does not obey any conservation law. Therefore, almost by construction, we do not see the beautiful picture that emerges in \cite{Rakovszky17,Khemani17}. There the random quantum circuit obeys a conservation law, where the block diagonal random unitary preserves local $z$-spin, \ie is Haar-random within fixed charge sectors. In these models, they find an emergent coupled diffusion process, where non-conserved operators propagate ballistically, with fronts that spread diffusively, and where conserved charge dissipates as conserved operators decay to non-conserved operators. This coupled process gives rise to long hydrodynamic tails in the weight of the evolving operator. Instead, the models we consider here just exhibit ballistic growth with a diffusive spreading of the ballistic front. 

While operator growth in the unitary random circuits can be understood as a biased random walk, which gives rise to a diffusive front, in the symmetric circuits we find that the endpoint dynamics evolve as different random walks. The endpoints of evolving strings in the COE and CSE random circuits are governed by a persistent random walk with correlation between steps. The edge of an evolving Pauli string in the $O(d)$ and $Sp(d)$ random circuits is a correlated random walk with an internal state. In each case we derive the butterfly velocity and diffusion constant of the evolving operator by solving the random walk. For all the models, we find that adding symmetry slows down the ballistic growth. 

We should view these constructions as models of local chaotic dynamics with antiunitary symmetries. Physical systems realizing time-reversal symmetry evolve in quotients of the unitary group \cite{DysonSym}. Dyson's insight was that systems which realize or break antiunitary symmetries fall into three symmetry classes. For those realizing time-reversal symmetry, the unitary time-evolution operator $e^{iHt}$ is valued in $U(d)/O(d)$ and $U(d)/Sp(d)$, compact symmetric spaces which can be realized as subsets of the unitary group. The COE and CSE are the unitary matrix ensembles for these two symmetric spaces. 

We will start by reviewing operator growth in unitary random quantum circuits. We then introduce four classes of symmetric random circuits and derive the respective transition probabilities of the stochastic process evolving Pauli strings. Then we discuss operator growth in the symmetric random circuits and derive the butterfly velocity and diffusion constant by solving the fictitious random walk governing operator growth. 

\subsubsection*{Random quantum circuits}
\begin{figure}
\centering
\begin{tikzpicture}[thick,scale=0.55]
\node[anchor=south,inner sep=0] at (0,0) {\includegraphics[width=0.45\linewidth]{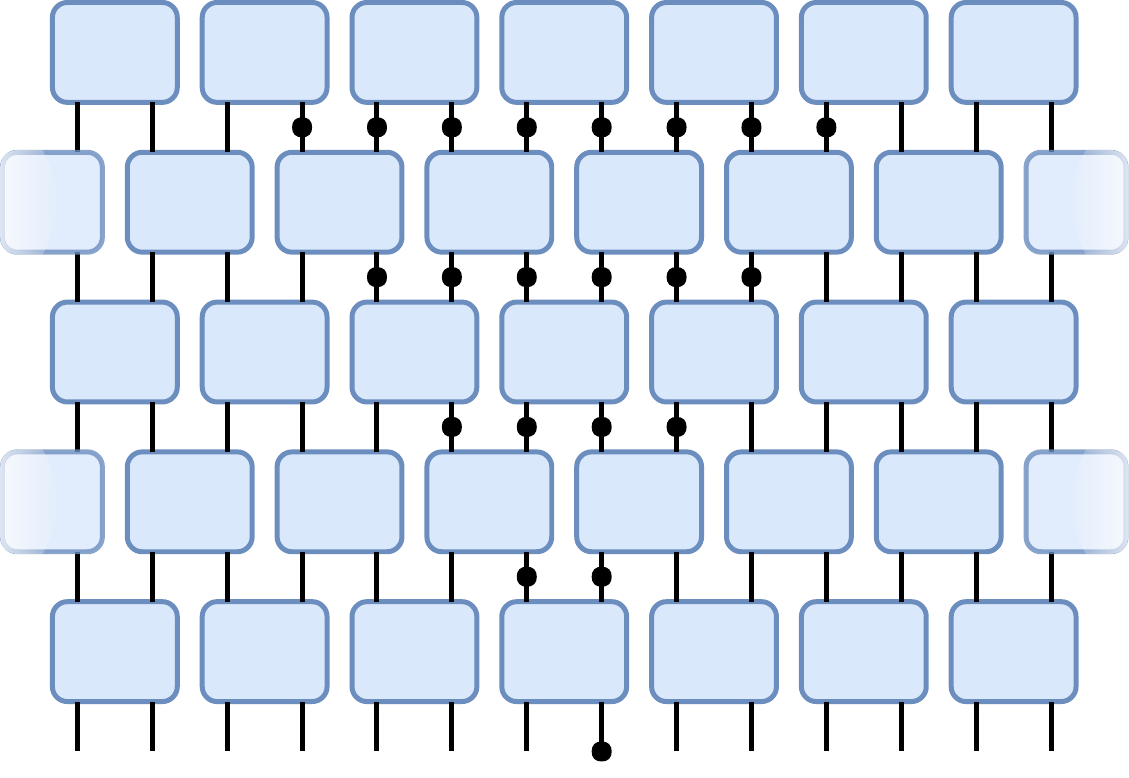}};
\draw[thick,->] (0.12,0.2) -- (4,8.8);
\draw[thick,->] (-0.12,0.2) -- (-4,8.8);
\end{tikzpicture}
\begin{tikzpicture}[thick,scale=0.6,baseline=-0.4cm]
\draw[thick,->] (0,0) -- (0,6) node[anchor=south] {$t$};
\end{tikzpicture}
\caption{Random quantum circuits built from staggered layers of 2-site unitaries, where each unitary is drawn at random from the unitary group or a subgroup or subset thereof, give rise to the ballistic growth of the support of an initially local operator.}
\label{fig:RQCdiag}
\end{figure}
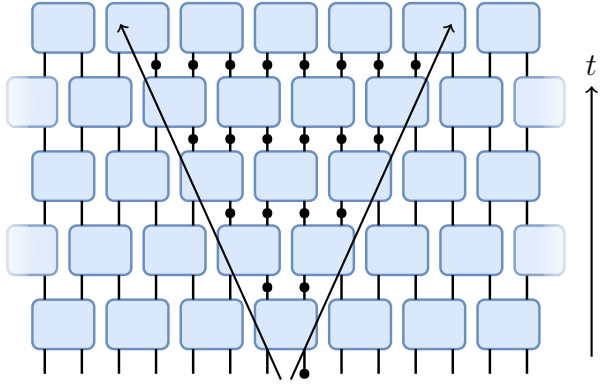

Consider a one-dimensional chain of $n$ qudits, with local dimension $q$, evolved by a random circuit built from layers of 2-site unitaries. Each layer or time-step of the random circuit alternates between acting on all even links between qudits at even time-steps, and odd links at odd time steps. Explicitly, the $t$-th layer of the circuit for even $t$, is given by the tensor product of 2-site unitaries $U_{i,i+1}$ for even $i$, and the $t$-th layer for odd $t$ by tensoring $U_{i,i+1}$ for odd $i$. We denote the unitary implementing the $t$-th layer as $U_t$, and the evolution to time $t$ is simply the product of $t$ layers of the circuit. The architecture of the random quantum circuit is shown in Fig.~\ref{fig:RQCdiag}. Each 2-site unitary $U_{i,i+1}$ is drawn at random from the unitary group $U(q^2)$ or from some subgroup or quotient thereof. We define the dimension of a 2-site unitary as $d\equiv q^2$, so that the local unitaries are drawn from $U(d)$.
%(and $U(4)$ in the case of qubits). 

%First, we give a general overview of operator growth in random circuits, which applies to all classes of random circuits in this note.
We'll go over the basic structure of operator growth in random circuits to set the stage for the rest of the discussion. The discussion here is general and applies to all classes of random circuits in this note. Recall that we can expand any operator in a basis of operators (like Pauli strings for qubit systems) as
\begin{equation}
\op(t) = \sum_p \gamma_p(t) \op_p\,.
\end{equation}
We will refer to the elements in the basis as Pauli strings regardless of local dimension; the only difference is that generalized Paulis for qudits are no longer necessarily Hermitian, but this does not affect the discussion. We should think about the coefficients $\gamma_p(t)$ as probabilities of finding the operator $\op_p$ in the growing operator, or equivalently the weight of the growing operator on a given Pauli string. Unitary evolution and the orthonormality of Paulis $\frac{1}{q^{2n}}\Tr (\op_a^\dagger \op_b) = \delta_{ab}$ means the operator norm is conserved under time-evolution, and thus the probabilities are conserved $\sum_p |\gamma_p(t)|^2 = 1$.

Consider the evolution of a local operator $\op_0$ in the random circuit. The local 2-site unitaries will act on the operator and the range of its support will grow ballistically, spreading outwards by one site at each side every time step, as shown in Fig.~\ref{fig:RQCdiag}. While the operator will grow to a linear combination of all Pauli strings with support on the $2t+1$ sites at time $t$ in the light-cone of the operator, we want to know the distribution on those Pauli strings, \ie the shape of the support of that operator on Pauli strings. We can define the weight of the evolving operator on its left/right edge as
\cite{Nahum17,vonKey17}
\begin{equation}
\rho_{\rm L/R}(x,t) = \sum_{p\in \op_{\rm L/R}(x)}  |\gamma_p(t)|^2\,,
\end{equation}
where we sum over all Pauli strings with the left or right-most non-identity Pauli operator at site $x$. The edges of the growing operator grow ballistically but also spread diffusively, such that $\rho_{\rm L/R}(x,t)$ obeys a simple biased diffusion equation \cite{Nahum17,vonKey17}
\begin{equation}
\p_t \rho(x,t) = v_B \p_x \rho(x,t) + D\p^2_x \rho(x,t) \with \rho(x,t) = \frac{1}{\sqrt{4\pi D t}} e^{- (x-v_B t)^2/4Dt}\,.
\label{eq:bde}
\end{equation}

We can solve for this distribution of the support of the operator by studying the random walking of the edges of the Pauli strings. The random circuits define a Marokv chain governing the internal dynamics of the growing operator, where each Pauli string in the growing operator is stochastically updated to another Pauli string. As we will discuss, each local 2-site unitary updates two-site pairs of the Pauli string, taking identities to themselves and non-identity Paulis to other non-identity Paulis. For instance, two sites in a Pauli string might get updated like $(XY) \ra (ZX)$. An update like this occurs for each two sites, \ie at every link, alternating between even and odd links in the circuit. Note that it is convenient to define our coordinate system on the links, \ie with respect to the 2-site gates, instead of the qudits. This way, the edge of an evolving Pauli either moves forwards or backwards at each time step.

\begin{figure}
\centering
\begin{tikzpicture}[thick,scale=0.5,baseline=-1cm]
\node[anchor=south,inner sep=0] at (0,0) {\includegraphics[width=0.20\linewidth]{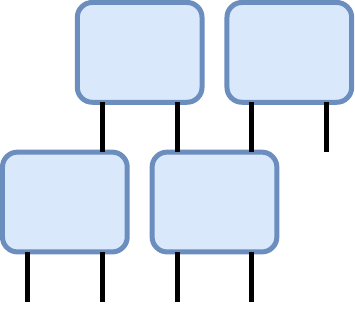}};
\draw[line width=1.3pt,->] (0.6,1.9) -- (-0.9,4.9);
\node at (0,-0.3) {$X$};
\node at (1.4,-0.3) {$\iden$};
\node at (-2.8,-0.3) {$Y$};
\node at (-1.4,-0.3) {$Z$};
\node at (2.8,2.55) {$\iden$};
\node at (1.7,3.4) {$\iden$};
\node at (-0.95,3.4) {$X$};
\node at (0.4,3.4) {$Z$};
\node at (-4,2) {$\cdots$};
\node at (-2.5,4.7) {$\cdots$};
\end{tikzpicture}
\hspace*{1cm}
\includegraphics[width=0.45\linewidth]{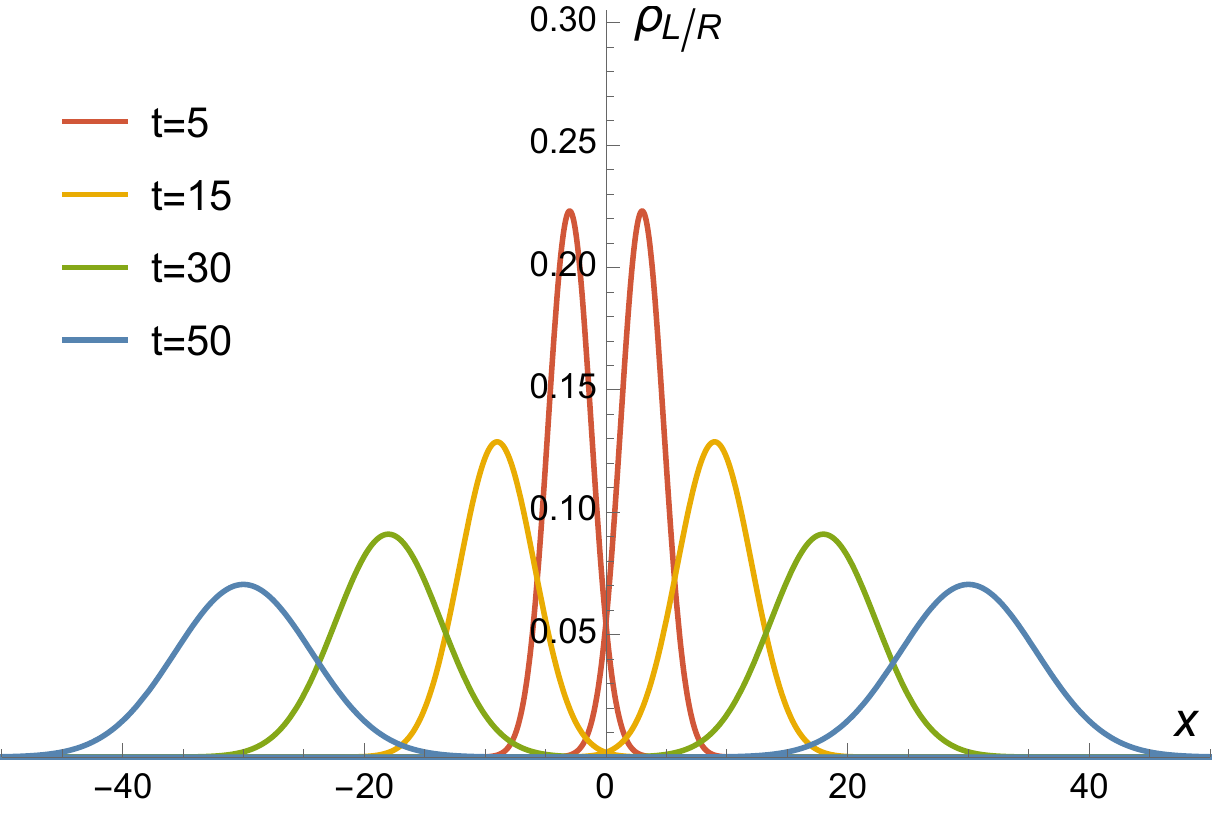}
\caption{Operator growth in random quantum circuits. On the left: an example of the right edge of a Pauli string moving back after evolution by a layer of the circuit. On the right: the evolution of the weight $\rho_{\rm L/R}(x,t)$ of the growing operator on its left/right edge.
}
\label{fig:RQCback}
\end{figure}
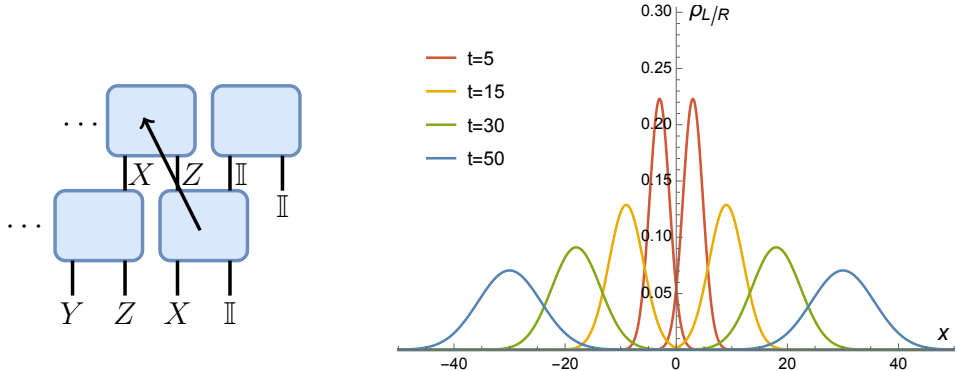

In the random circuit evolution, each Pauli string is random walking through the space of Pauli strings. Moreover, the end of each Pauli string has some probability of moving back. For instance, say the two Pauli operator at the right-most gate is $(X\,\iden)$. There is some probability that this gets updated to $(Z\,\iden)$ and the gate to the right in the following time-step only acts on identities. Thus in the coordinate system on the links, the operator moves back, as shown in Fig.~\ref{fig:RQCback}. This is a random walk of the edges of the Pauli strings making up the operator. The biased random walk governing the edge of an evolving operator gives rise to a biased diffusion process at long times, where the weight of the operator on the left/right edges obeys a biased diffusion equation in Eq.~\ref{eq:bde}.

\section{Operator growth in unitary random circuits}\label{sec:Urqc}
We start by reviewing the story for unitary random circuits, where each local 2-site unitary in Fig.~\ref{fig:RQCdiag} is drawn Haar-randomly from $U(d)$ with $d\equiv q^2$. This closely follows the analysis in \cite{HarrowLow08,LowThesis,Brown12,Nahum17}. 
Consider the evolution of some local operator $\op(t)$, acting on a single site at time zero. We can solve for the growth of this operator by solving for the coefficients $|\gamma_p(t)|^2$
\begin{equation}
|\gamma_p(t)|^2 = \frac{1}{q^{2n}} \big| \Tr\big(\op(t) \op_p\big)\big|^2 =  \frac{1}{q^{2n}} \Tr\big(U_t \op(t-1) U_t^\dagger \op_p\big) \Tr\big(U_t \op(t-1) U_t^\dagger \op_p\big)\,,
\end{equation}
where $U_t$ is the layer implementing the $t$-th time step acting on the operator at $t-1$. Expanding in the operator basis
\begin{equation}
|\gamma_p(t)|^2 = \frac{1}{q^{2n}} \sum_{a,b} \gamma_a(t-1) \gamma_b(t-1) \Tr\big(U_t \op_a U_t^\dagger \op_p\big) \Tr\big(U_t \op_b U_t^\dagger \op_p\big)\,,
\label{eq:opcoeffsU}
\end{equation}
we now Haar average the expression, decomposing the expression into a product over 2-site operators and 2-site random unitaries. We review Haar integration in App.~\ref{app:Haar}. For a single 2-site operator acted on by a 2-site gate, the expression we find averaging using the 2nd moment is
\begin{equation}
\int_{U(d)} dU\, \Tr\big(U \op_a U^\dagger \op_p\big) \Tr\big(U \op_b U^\dagger \op_p\big) = \frac{d^2}{d^2-1} \delta_{a,b} \big(d^2 \delta_{a,1} \delta_{p,1} + 1 - \delta_{a,1} - \delta_{p,1} \big)\,,
\end{equation}
where $\delta_{p,1} = 1$ if $\op_p$ is the identity. 
%Proceeding, we have
%\begin{equation}
%|\gamma_p(t)|^2 = \sum_{a} |\gamma_a(t-1)|^2 \prod_s \Big( \delta^{(s)}_{a,1} \delta^{(s)}_{p,1} + \frac{1}{d^2-1} (1-\delta^{(s)}_{a,1})(1 - \delta^{(s)}_{p,1})\Big)\,,
%\end{equation}
Taking the product over all pairs of sites on which the 2-site random unitaries act, we can write Eq.~\eqref{eq:opcoeffsU} as
\begin{equation}
|\gamma_p(t)|^2 = \sum_{a} S_{pa} |\gamma_a(t-1)|^2\,, \where  S_{pa} = \prod_s \Big( \delta^{(s)}_{a,1} \delta^{(s)}_{p,1} + \frac{1}{d^2-1} (1-\delta^{(s)}_{a,1})(1 - \delta^{(s)}_{p,1})\Big)\,,
\end{equation}
which determines the growth of the operator at each time step. The symmetric matrix $ S_{pa}$ is the transition matrix of a Markov chain on Pauli strings, where the Pauli string is updated stocastically at pairs of sites with transition probabilities determined by $S_{pa}$. The matrix $S_{pa}$ essentially tells us that at each pair of sites we take $\iden\,\iden \ra \iden\,\iden $ or a non-identity 2-site Pauli operator to any of the non-identity Paulis $\op_p \ra \op_a$ each with prob $1/(d^2-1)$, which makes sense as there are $d^2-1$ non-identity 2-site Paulis.

\subsubsection*{Unitary random walk}
The transition probabilities for the Markov chain on 2-site Pauli operators, given by a Haar-random 2-site unitary, are
\begin{align}
\begin{aligned}
\iden\,\iden &\ra \iden\,\iden &\qquad &\text{with prob}\quad 1\nn
\op_p &\ra \op_a &\qquad &\text{with prob}\quad 1/(d^2-1)\,.
\end{aligned}
\end{align}
These local updates on 2-site pairs of operators in the Pauli string give rise to a biased diffusion process. 
At the left and right-most gates, the update rule takes the a 2-site operator $\op_p$ to another non-identity 2-site operator. But the operator moves back if an $\iden$ appears at the end of the string. As there are $q^4-1$ non-identity 2-site Paulis and $q^2-1$ of them have an identity at the outer site, the probability of the edge of the operator moving back is $p = (q^2-1)/(q^4-1)$. These transition probabilities define a biased random walk for the edge of an evolving operator with $v_B = 1-2p$ and $D=2p(1-p)$. We briefly review random walks in App.~\ref{app:RW}. At late times, the weight of the operator on the left/right edges $\rho_{L/R}(x,t)$ obeys a biased diffusion equation \cite{Nahum17,vonKey17}, with butterfly velocity and diffusion constant 
\begin{equation}
v_B = \frac{q^2-1}{q^2+1} \and D = \frac{2 q^2}{(q^2+1)^2}\,.
\label{eq:Uvb}
\end{equation}

In the limit of large local dimension $q\ra \infty$, we have $v_B \ra 1$, achieving the light-cone velocity, and $D \ra 0$, so the operator moves forward deterministically. The butterfly velocity and diffusion constant in Eq.~\eqref{eq:Uvb} have the following $1/q$ expansions
\begin{equation}
v_B \approx 1 - \frac{2}{q^2} + \frac{2}{q^4} - \frac{2}{q^6} \quad\and\quad D \approx \frac{2}{q^2} - \frac{4}{q^4} + \frac{6}{q^6}\,.
\label{eq:Uvbexp}
\end{equation}

\subsubsection*{OTOCs for random unitary circuits}
The out-of-time ordered correlation function (OTOC) is a 4-point function of a pair of operators evaluated in (thermal) states, $\vev{A B(t) A B(t)}$. The quantity has been understood to probe chaotic dynamics in quantum many-body systems and gives a precise sense \cite{MSSbound} in which many-body systems, including black holes and holographic systems \cite{SSbutterfly,SSstringy}, are fast scramblers. The OTO commutator $\vev{|[A,B(t)]|^2}$ measures the failure of $B(t)$ to commute with $A$, where two of its constituent 4-point functions are OTOCs; the exponential growth of an operator corresponds to the decay of the OTOC \cite{LocalizedShocks}.\footnote{Using quantum information theoretic ideas, OTOCs have been understood as quantifying scrambling and randomness \cite{ChaosChannels,ChaosDesign,ChaosRMT}.}

Solving the operator growth in unitary circuits allows one to estimate the OTO commutator\footnote{ A universal form for the growth of the OTO commutator was proposed in \cite{SwingleMPS,KhemaniVel18}.} in different regimes of the passing front as the operator $B(t)$ spreads to $A$. Outside the lightcone of $B$, the commutator with $A$ must be zero. There is an early-time growth, with $A$ inside the lightcone of $B$ but outside the $\sqrt{t}$ front, with a distance-dependent analog of a Lyapunov growth \cite{vonKey17}. Within the diffusive front, the OTO commutator grows to an order one value as an integrated Gaussian, and afterwards relaxes to its asymptotic value. The symmetric circuit models considered in this work exhibit the same OTO commutator growth as in the unitary circuits, simply with different velocity and diffusion constants for the front.

\section{Operator growth in symmetric random circuits}\label{sec:symRQC}
The random unitary circuit picture is nice, we see the emergence of a diffusive phenomenon from unitary dynamics. But as the random circuit breaks all symmetries and obeys no conservation laws, we also want to know what happens when we start building in symmetry in different ways. Consider random circuits constructed out of local gates randomly drawn from ensembles with symmetry. We will consider four different symmetry classes, local orthogonal and symplectic matrices, and the two quotients of the unitary group $U(d)/O(d)$ and $U(d)/Sp(d)$, corresponding to the COE and CSE. In each case the analysis is more complicated than the analysis for the random unitary circuits, but we can in fact solve the operator growth analytically. In the unitary random circuits, the evolution of a Pauli string is a Markov chain and the dynamics of the end of the operator is itself an autonomous Markov process. In the symmetric circuits the evolution is also a Markov chain on Pauli strings, in each case with different transition probabilities. But the dynamics of the endpoints are no longer Markovian, as there are correlations between different time steps. This is a persistent random walk, where the persistence refers to the operator having some `inertia,' \ie wanting to move in the same direction as the previous time step. In our symmetric random circuits, we find that the successive time steps are anticorrelated, thus the random walks are biased and anti-persistent. 

In each of the four symmetric random circuits we consider, we compute the probability of an operator moving back, which gives the butterfly velocity, but the analysis of the diffusion constant is a little more subtle. The anticorrelation reduces the diffusion constant from its uncorrelated value, \ie the one that would be computed with $p$ in the uncorrelated random walk. In the COE and CSE random walks, we can compute the diffusion constants directly, but for the $O(d)$ and $Sp(d)$ random walks we need to derive the diffusion equation.

These circuit constructions are a simple application of the tools developed in \cite{ChaosSym,NHJthesis}, making use of Weingarten calculus for compact Lie groups and compact symmetric spaces. We give a quick overview of Haar integrals in App.~\ref{app:Haar}, but Weingarten calculus for Lie groups and compact symmetric spaces was reviewed in more detail in \cite{NHJthesis}. The methods for performing Haar averages over compact Lie groups were worked out in \cite{Collins02,Collins04,CollinsMat09}, and extended to compact symmetric spaces in \cite{Matsumoto11,Matsumoto13}. In computing the random circuit averages, we have relied heavily on the methods developed there.

Before discussing our random circuit models, we summarize our results in the table below
\begin{center}
\begin{tabular}{ccll}
{Random circuit model}& $v_B$& $v_B$ for $q=2$& $v_B$ at large $q$ \\ \hline \vspace*{-6pt}\\
Unitary $U(d)$& $\frac{q^2-1}{q^2+1}$& ~$\frac{3}{5} ~\quad (=0.6)$& ~$1 - \frac{2}{q^2} + \frac{2}{q^4} - \frac{2}{q^6}$ \vspace*{4pt}\\
COE $U(d)/O(d)$& \eqref{eq:vbCOE}& ~$\frac{153}{305} ~~ (\approx 0.502)$& ~$1 - \frac{2}{q^2} - \frac{2}{q^4} + \frac{14}{q^6}$  \vspace*{4pt}\\
CSE $U(d)/Sp(d)$& \eqref{eq:vbCSE}& ~$\frac{11}{41} \quad (\approx 0.268)$& ~$1 - \frac{2}{q^2} - \frac{2}{q^4} - \frac{2}{q^6}$\vspace*{4pt}\\
Orthogonal $O(d)$& \eqref{eq:vbO}& ~$\frac{23}{39} \quad (\approx 0.589)$& ~$1 - \frac{2}{q^2} + \frac{6}{q^5} - \frac{4}{q^6}$ \vspace*{4pt}\\
Symplectic $Sp(d)$& \eqref{eq:vbSp}& ~$\frac{7}{15} \quad (\approx0.467)$ & ~$1 - \frac{2}{q^2} - \frac{2}{q^5} + \frac{4}{q^7}$ \vspace*{4pt}\\
\end{tabular}
\end{center}

We find that the symmetric random circuits are always slower than the unitary circuits; adding symmetry slows down operator growth. For qubits, the unitary circuits have $v_B = 3/5$. In the time-reversal symmetric circuits, the naive butterfly velocity is $v_B=1/2$ (COE) and $v_B=1/4$ (CSE), but there is a subtle (yet computable) effect which increases the value slightly. For $O(d)$ qubit circuits, $v_B = 23/39$ and for $Sp(d)$ circuits, $v_B=7/15$. 

\begin{figure}[htb!]
\centering
\includegraphics[width=0.5\linewidth]{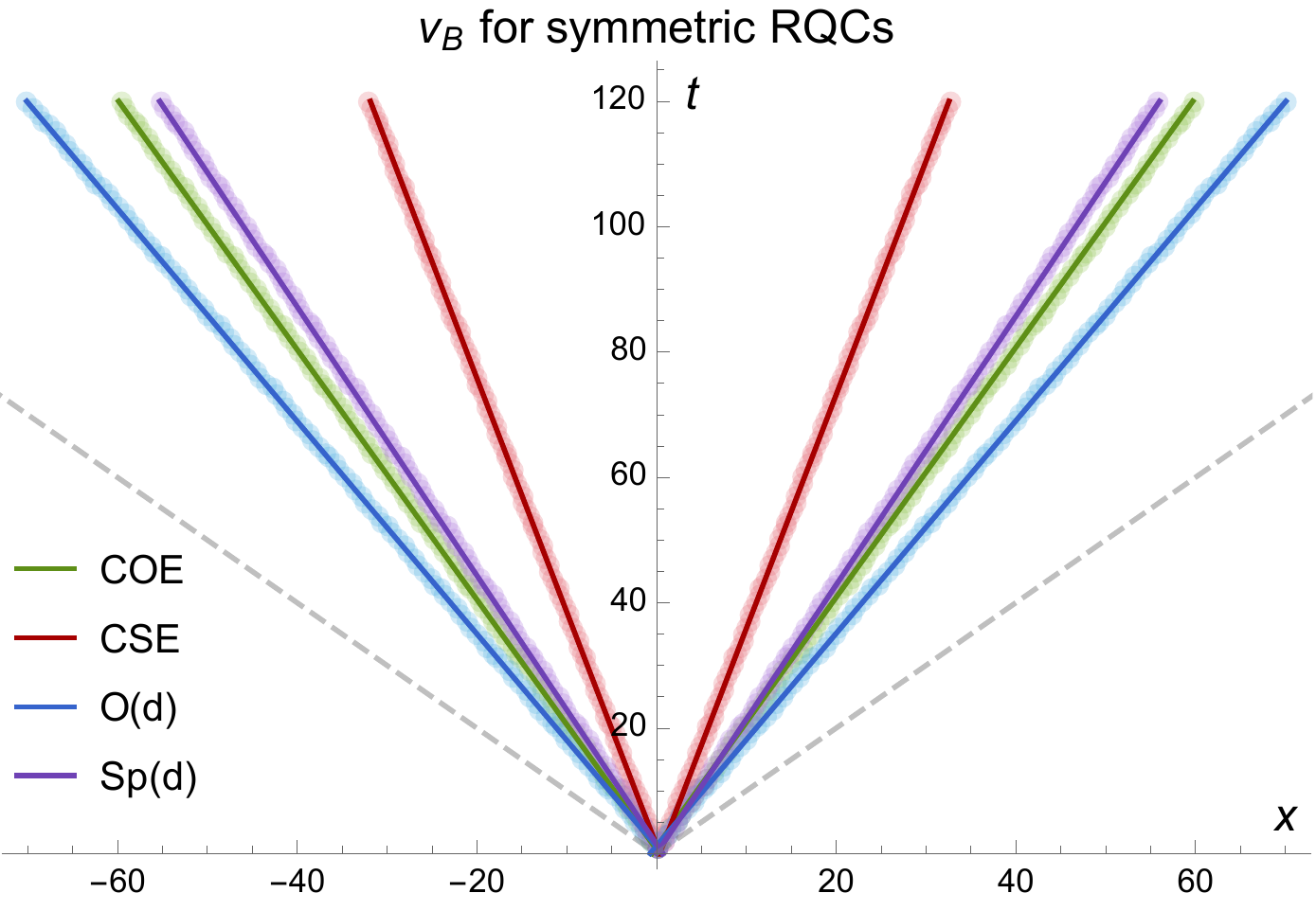}
\caption{Butterfly velocities in the random symmetric circuits with local qubits. The numerical simulation of growing operators (lighter colored points) agrees well with the analytic expressions (darker lines) for $v_B$ in each class of circuits. The gray dotted line is the lightcone velocity.
}
\label{fig:RQCvb}
\end{figure}

In Fig.~\ref{fig:RQCvb}, we provide some numerical checks of the results derived in the table above. We simulated operator growth in the four classes of symmetric circuits with local qubits, averaged over an ensemble size of 20,000. The numerical fit of the growth of the edges of the operator agrees very well with all values of the $q=2$ velocities in the table above. We plot the numerics alongside the analytic expressions for $v_B$ using the constant offsets given by the fit. We also numerically check the diffusion constants and again find good agreement.

\subsection{Operator growth in random COE circuits}\label{sec:COErqc}
Consider now a circuit of 2-site random unitaries drawn from the circular orthogonal ensemble (COE), \ie random elements of the symmetric space $U(d)/O(d)$.  
To understand the operator growth, we again consider the action of a single random gate
\begin{equation}
\int_{\rm COE} dU\, \Tr\big(U \op_a U^\dagger \op_p\big) \Tr\big(U \op_b U^\dagger \op_p\big)\,.
\end{equation}
We can compute the average using the second moment of the $U(d)/O(d)$ \cite{Matsumoto11,Matsumoto13}, as is briefly review in App.~\ref{app:Haar}, and find the transition matrix for 2-site Pauli operators
\begin{align}
S_{pa} = \prod_s\Big( \delta^{(s)}_{a,1} \delta^{(s)}_{p,1} + \frac{1}{d(d+1)(d+3)} \big( &(d+4)(\delta^{(s)}_{a,1}-1)(\delta^{(s)}_{p,1}-1)\nn 
&+ 2(d+2) \delta^{(s)}_{a,p} (1-\delta^{(s)}_{p,1}) +\frac{2}{d} (1-(-1)^{(a,p)}\big) \Big)\,,
\end{align}
where $(-1)^{(a,p)}$ is $+1$ if the Paulis commute $[ \op_p,\op_a] = 0$, and $-1$ if they anticommute $\{\op_p,\op_a\} = 0$. More precisely, Pauli operators commute if their symplectic inner product $(a,p)$ equals zero. This is a little more involved than the unitary case; here we see that the identity is mapped to the identity, and non-identity Paulis are taken to non-identity Paulis, but with a higher probability of being mapped to itself, and different probabilities for Paulis they commute or anticommute with.

From the above transition matrix, we can list the transition probabilities for 2-site Pauli operators
\begin{align*}
\iden\,\iden &\ra \iden\,\iden && \text{with prob} \quad 1 \\
\op_p &\ra \op_{a=p} &&\text{with prob}\quad p_s = (3d+8)/d(d+1)(d+3)\\
\op_p &\ra \op_{a \neq p} \and [\op_a,\op_p] = 0 &&\text{with prob}\quad p_c = (d+4)/d(d+1)(d+3)\\
\op_p &\ra \op_{a \neq p} \and \{\op_a,\op_p\} = 0 &&\text{with prob}\quad p_a = (d+2)^2/d^2(d+1)(d+3) \,.
\end{align*}
As a sanity check, recall that exactly half of Pauli operators commute with a Pauli operator $\op_p$ (including itself and the identity) and half anti-commute. So we see that the probabilities for a non-identity Pauli operator add to one: $p_s + p_c(d^2/2-2) + p_ad^2/2 = 1$.

\subsubsection*{COE random walk}
Now we study operator growth in COE random circuits. The fact that an operator is more likely to update to itself already indicates the the butterfly velocity will be slower. The endpoint of a Pauli string is a random walk with correlation between the timesteps, as the motion of the random walker depends on the action at the previous time step. Thus, the random walk itself is non-Markovian, but the process of updates on the endpoint is a two-state Markov chain, with probabilities of moving forward and backward depending on whether the random walker moved forward or backward at the previous timestep. 

Let $p_1$ be the probability the operator moves back after moving forward and let $p_2$ be the probability the operator moves back after moving back. The probability that an operator moves back at time $t$, $p(t)$, satisfies the evolution equation
\begin{equation}
p(t) = (p_2 - p_1)p(t-1) + p_1\,.
\end{equation}
Solving for the stationary probability or summing the series, we can solve for $p$ and find
\begin{equation}
p = \frac{p_1}{1+p_1-p_2}\,.
\label{eq:stprob}
\end{equation}
For an operator that has moved forward at the previous timestep, \eg we have $X\iden$ at the rightmost gate, the probability the string moves back is $p_1 = p_s+(d/2-2)p_c+(d/2) p_a$ (the probability it goes to itself plus the probability of going to commuting or anticommuting operators that also move back). If the operator has moved back at a previous timestep, then the operator at the edge of the string has a non-identity Pauli at the rightmost site, \eg $XZ$ and the rightmost gate, and will move forward with higher probability (as updating to itself moves the operator forward). In this case the probability the operator moves back is naively $p_2 = (d/2 -1) p_c + (d/2) p_a$. 

Solving for $p$ and $v_B$ using the COE transition probabilities we find that, when $q=2$
\begin{equation}
\text{For qubits}:\quad p = 1/4 \and v_B=1/2\,.
\end{equation}

But there is a subtlety which modifies $p_2$ and increases the butterfly velocity slightly. The probability of moving back twice, $p_2$, depends on the operator coming from inside the circuit. Any non-identity Pauli will commute/anticommute with half of the 1-site operators that move us back, but if $\iden$ appears at the left site of the gate, it will commute with all operator that move us back. Assuming that the interior of the circuit is rapidly mixing and operators appear with uniform probability, we have $p_2 = \frac{d-1}{d}\big((d/2 -1) p_c + (d/2) p_a\big) + \frac{1}{d}(d-1)p_c$ .

%if the operator at the right most gate is $XZ$, with $Z$ coming from the operator moving back and $X$ coming from the second right-most gate. The operators that move us back are $\{X\iden,Y\iden,Z\iden\}$, with which $XZ$ anticommutes with two and commutes with one. But if the operator coming from the second rightmost gate is the identity, \eg $\iden Z$, then we commute with all operators moving back.

We then find the probability of a Pauli string moving back in the COE circuit in terms of the local dimension $q$
\begin{equation}
p = \frac{q^2 (q^4+5 q^2+2)}{(q^2+1) (q^6+3 q^4+2 q^2+2)}\,,
\end{equation}
from which we find the butterfly velocity for COE random circuits $v_B = 1-2p$ to be
\begin{equation}
v_B = \frac{(q^2-1)^2 (q^4+4 q^2+2)}{(q^2+1) (q^6+3 q^4+2 q^2+2)}\,.
\label{eq:vbCOE}
\end{equation}
For qubits, this increases $v_B $ from $1/2$ to $v_B= 153/305\approx 0.502$, which is somewhat slower than unitary circuits with $v_B = 3/5$. The above expression is exact for any local dimension $q$, but we can expand in powers of $q$ to find
\begin{equation}
v_B = 1 - \frac{2}{q^2} - \frac{2}{q^4} + \frac{14}{q^6} + \ldots
\end{equation}
for the COE circuits. 

\subsubsection*{Diffusion in random COE circuits}
Recall that for the growing operator in the random COE circuits, we derived the probability of an operator moving back to be
\begin{equation}
p(t) = p_1 + p(t-1) (p_2-p_1)\,,
\label{eq:COEprob}
\end{equation}
with the probabilities $p_1$ and $p_2$ defined above. The dependence on the probability of moving back introduces correlation between the time steps. The evolution of the end of the operator in the random circuit is a persistent biased random walk, where there is a direct correlation with the motion at a previous timestep. Persistent random walks \cite{Taylor22,GoldsteinDiff,RenshawCorr,WeissRW} were introduced to accommodate inertial effects in Brownian motion. 
As the prefactor $p_2-p_1$ is negative, there is an anticorrelation; the COE random walker is more inclined to move in the opposite direction from its previous motion. 

Recall that the position of the random walker $X(t) = \sum_i^t x_i$ is given as the sum of increments $x_i=\pm 1$, as reviewed in App.~\ref{app:RW}. In the persistent random walk, the correlation between timesteps does not affect the mean $\vev{X(t)}$, and computing the stationary probability in Eq.~\eqref{eq:stprob} directly gives us the butterfly velocity $v_B$. But correlations between timesteps affect the variance $\vev{X(t)}_c$ and thus the diffusion constant. Generally, we have
\begin{equation}
\vev{X(t)^2}_c = \sum_{i,j} \vev{x_i x_j}_c = 4tp(1-p) + \sum_{i\neq j} \vev{x_i x_j}_c\,,
\label{eq:Xvar}
\end{equation}
where $\vev{\cdot}_c$ is the connected correlator. In first term we find the uncorrelated diffusion constant $D_0\equiv 2p(1-p)$. Define $c(|i-j|) \equiv \vev{x_i x_j}$ as the correlator between time steps, where we have translational invariance in time. Clearly, $c(0) = 4tp(1-p) = 2D_0 t$. To compute the correlations for the COE random walker, we need to look closer at the direct correlation between time steps.

The probability of moving back in an COE circuit, Eq.~\eqref{eq:COEprob}, means that the random variable $x_i$ at a time step $i$, is a sum of two random variables
\begin{equation}
x_i =  \alpha x_{i-1} + p_1 x'_i\,, \where \alpha = (p_2-p_1)
\label{eq:xrv}
\end{equation}
and where $x_{i-1}$ is the step taken at the previous timestep and $x'_i$ is an iid random variable, with mean and variance to be determined. There is now a direct correlation between timesteps given by the coefficient of the $x_{i-1}$ variable $\alpha=p_2-p_1$.  We already know the mean and variance of $x_i$ to be $\vev{x_i} = (1-2p)$  and $\vev{x_i^2}_c = 4p(1-p)$. We can fix the mean and variance of the random variable $x_i'$ from Eq.~\eqref{eq:xrv}, and find
\begin{equation}
\vev{x_i'} = \frac{1-2p}{p}\,, \and \vev{x_i'^2}_c = 4(1-p)\frac{1-p_1+p_2}{p_1}\,.
\label{eq:xrvm}
\end{equation}
First we note that if $p_2=p_1$, the correlation between time-steps vanishes and the endpoint dynamics are uncorrelated and Markovian. In this case the probability of moving back is $p=p_1$. We see that in terms of $x_i'$ the mean and variance of $x_i$ from Eq.~\eqref{eq:xrvm} become $\vev{x_i} = (1-2p_1)$ and $\vev{x_i^2}_c = 4p_1(1-p_1)$, as expected. As a second sanity check, if $p_1=0$, the probability of moving is zero and the endpoint deterministically moves forward. 

We can now compute the correlator between random steps at successive times and find that for a difference 
\begin{equation}
c(t) = \vev{x_i x_j}_c = 2D_0\alpha^{t}\,, \where t = |i-j|\,.
\end{equation}
Therefore, rewriting the sum in Eq.~\eqref{eq:Xvar} by making a change of variables and then summing the resulting series, we find
\begin{equation}
\vev{X(t)^2}_c = 2D_0t + 2\sum_{t'=1}^t (t-t') c(t') = 2D_0t \frac{1+\alpha}{1-\alpha} -\frac{\alpha  (1-\alpha ^t)}{(1-\alpha )^2}\,.
\end{equation}
For times $t\gg 1$, we have
\begin{equation}
\vev{X(t)^2}_c \approx 2D t\,, \where D = \frac{1+\alpha}{1-\alpha} D_0\,,
\end{equation}
where $D_0 = 2p(1-p)$ is the uncorrelated diffusion constant and $\alpha =p_2-p_1$ is the persistence of the random walker. For qubits, we find $D \approx 0.31$. 

This was more of a Langevin derivation of diffusion, but we can arrive at the same result employing more of a Fokker-Planck approach, explicitly deriving the diffusion equation. In App.~\ref{app:diff}, we derive the same diffusion constant from the difference equations for the 2-state Markov chain governing the motion of the COE random walk.

\subsection{Operator growth in random CSE circuits}\label{sec:CSErqc}
Consider now a random quantum circuit with 2-site unitaries drawn at random from the circular symplectic ensemble (CSE), corresponding to random instances of the symmetric space $U(d)/Sp(d)$.\footnote{It is common to denote the unitary symplectic group with even dimension $2d$, \ie defined as the intersection of symplectic matrices $Sp(2d,\C)$ and the unitary group. Here, for consistency of the discussion and ease in comparing formulae, we denote the symplectic group $Sp(d)$ and the compact symmetric space AII as $U(d)/Sp(d)$, keeping in mind that $d$ must be taken to be even in these two cases.}
To understand the operator growth, we again consider the action of a single random gate
\begin{equation}
\int_{\rm CSE} dU\, \Tr\big(U \op_a U^\dagger \op_p\big) \Tr\big(U \op_b U^\dagger \op_p\big)\,,
\end{equation}
which we compute using the second moment of the CSE \cite{Matsumoto13} and find the transition matrix
\begin{align}
S_{pa} = \prod_s\Big( \delta^{(s)}_{a,1} \delta^{(s)}_{p,1} + \frac{1}{d(d-3)(d-1)} \big( &(d-4)(\delta^{(s)}_{a,1}-1)(\delta^{(s)}_{p,1}-1)\nn 
&+ 2(d-2) \delta^{(s)}_{a,p} (1-\delta^{(s)}_{p,1}) + \frac{2}{d} (1-(-1)^{(a,p)}\big) \Big)\,.
\end{align}
Again, like in the COE circuit, we see that the identity is mapped to the identity, and non-identity Paulis are mapped to a linear combination of all other non-identity Paulis, but with a higher probability of being mapped to itself. 

From $S_{pa}$, we find the transition probabilities for 2-site Pauli operators are
\begin{align}
\iden\,\iden &\ra \iden\,\iden & &\text{with prob}\quad 1 \nn
\op_p &\ra \op_{a =p}  & &\text{with prob}\quad p_s = (3d-8)/d(d-3)(d-1) \nn
\op_p &\ra \op_{a\neq p} \quad [\op_a,\op_p]=0 \quad & &\text{with prob}\quad p_c = (d-4)/d(d-3)(d-1) \nn
\op_p &\ra \op_{a\neq p} \quad \{\op_a,\op_p\}=0 \quad & &\text{with prob}\quad p_a = (d-2)^2/d^2(d-3)(d-1) 
\end{align}

The analysis is much the same as the COE random circuits, where there is a higher probability of an operator going to itself, and separate probabilities of an operator updating to operators which commute or anticommute with that operator. One interesting difference here is that for $q=2$, local qubits, the updated probabilities become
\begin{equation}
\text{For qubits:}\qquad p_s=1/3\,, \qquad p_a = 1/12\,, \qquad p_c = 0\,.
\end{equation}
So the probability an operator updates to itself is fairly high, and no 2-qubit operator updates to an operator with which it commutes. This means that operator growth will happen fairly slowly as the operators at the end of the Pauli string, \eg of the form $X \iden$, have a much higher chance of updating to themselves, and thus moving back.

\subsubsection*{CSE random walk}
Just as in the COE random circuit, we can solve for the probability for an evolving operator to move back for general local dimension $q$
\begin{equation}
p= \frac{(q^4-1)(q^2-2)}{q^8-3 q^6+q^4+q^2-2}\,.
\end{equation}
The naive butterfly velocity is $v_B = 1/4$, but taking into account the identity contribution to $p_2$, we find $v_B = 11/41 \approx 0.268$, substantially slower than in the other random circuits. The butterfly velocity for random CSE circuits is given in terms of $q$ as
\begin{equation}
v_B = \frac{q^8-5 q^6+5 q^4+3 q^2-6}{q^8-3 q^6+q^4+q^2-2}\,.
\label{eq:vbCSE}
\end{equation}
Expanding the above expression at large $q$ we find
\begin{equation}
v_B \approx 1 - \frac{2}{q^2} - \frac{2}{q^4} - \frac{2}{q^6}\,,
\end{equation}
giving the same leading order term in $1/q$, but different subleading corrections. 

The diffusion constant is also given just as in the COE case, where the correlation between steps is $\alpha = p_2-p_1$ in terms of the $p_s$, $p_c$, and $p_a$ defined for the CSE random circuits. Meaning the endpoint dynamics of the CSE random walker are also a biased random walk with anticorrelated steps. The diffusion constant is $D = \frac{1+\alpha}{1-\alpha} D_0$, where $D_0 = 2p(1-p)$ is the uncorrelated diffusion constant in terms of the stationary probability of moving back in the CSE circuits $p$. For qubits we find $D\approx 0.22$.

\subsection{Operator growth in random orthogonal circuits}\label{sec:Orqc}
Consider a quantum circuit of 2-site Haar random orthogonal operators. For a single 2-site operator, consider
\begin{align}
&\int_{O(d)} dU\, \Tr\big(U \op_a U^\dagger \op_p\big) \Tr\big(U \op_b U^\dagger \op_p\big)\nn
&\qquad = \frac{1}{(d-1)(d+2)}\Big( (d+1) d^3 \Tr(\op_a)\Tr(\op_b)\Tr(\op_p)^2 -d^2\Tr(\op_p)^2 \Tr(\op_a \op_b^T) + \ldots \,.
\end{align}
%\begin{align}
%\qquad~= d^2 \delta_{a,b} \Big( \delta_{a,1} \delta_{p,1} + \frac{1}{(d-1)(d+2)} \big(  &(\delta_{a,1}-1)(\delta_{p,1}-1) + (\delta_{a,1}-(-1)^{Y_a})(\delta_{p,1}-(-1)^{Y_p})\nn
% &+ \frac{1}{d} (1-(-1)^{Y_a})(1-(-1)^{Y_p})\big) \Big)\,,
%\end{align}
The second moment of the orthogonal group gives an expression with nine terms, and after some reworking we find Pauli strings evolve as $|\gamma_p(t)|^2 = \sum_{a} S_{pa} |\gamma_a(t-1)|^2$ with
\begin{align}
S_{pa} = \prod_s \Big( \delta^{(s)}_{a,1} \delta^{(s)}_{p,1} + \frac{1}{(d-1)(d+2)} \big(  &(\delta^{(s)}_{a,1}-1)(\delta^{(s)}_{p,1}-1) + (\delta^{(s)}_{a,1}-(-1)^{Y_a})(\delta^{(s)}_{p,1}-(-1)^{Y_p})\nn &+ \frac{1}{d} (1-(-1)^{Y_a})(1-(-1)^{Y_p})\big) \Big)\,,
\end{align}
where $Y_p$ equals zero or one depending on whether the operator $\op_p$ is even or odd under transposition. This defines the transition matrix of the orthogonal Markov process on Pauli strings, encoding local update rules on 2-site operators. In the orthogonal circuits, the matrix $S_{pa}$ tells us that at each site we take identities to identites, even non-identity Paulis to even and odd non-identity Paulis to odd:
\begin{align}
\begin{aligned}
\iden\,\iden &\ra \iden\,\iden &\qquad &\text{with prob}\quad 1\\
{\rm even}~ \op_p &\ra {\rm even}~\op_a & &\text{with prob}\quad p_e = 2/((d-1)(d+2))\\
{\rm odd}~ \op_p &\ra {\rm odd}~\op_a & &\text{with prob}\quad p_o = 2/(d(d-1))\,,
\end{aligned}
\end{align}
where even and odd refer to the sign under transposition $\op_p^T=\pm \op_p$, which is the natural action of the orthogonal group. 
%This is not too surprising given what we understand about $O(d)$ and the action of Paulis under the group. We see an interesting conservation law encoded into the growing wavefront. The wavefront is growing diffusively, but there are two coupled diffusion processes, where even Paulis are mapped to even and odd Paulis are mapped to odd.

\subsubsection*{Orthogonal random walk}
We can now solve for the operator growth in the random orthogonal circuits. The probability of an operator moving back depends on whether the operator at the farthest most gate is even or odd, and further depends on whether the action at the previous timestep was a move forward or backward. The structure of the alternating gates at each time step means that even operators output from a gate can input odd operators at the next gate. Consider $YY$, an even 2-site operator, if this operator appears at the edge of the growing operator, then the input to the gate of the farthest gate at the next time step is $Y\iden$, an odd operator. So we need to treat the dynamics of even and odd operators carefully.

We start by discussing the case for qubits and then derive formulae for general local dimension $q$. The probability of an even Pauli moving back is $2/9$, \ie if we get a $X\iden$ or $Z\iden$. For odd Paulis we have $1/6$, \ie if we generate a $Y\iden$. Denote the probabilities of even and odd operators moving forwards or backwards as 
\begin{equation}
p_{\leftarrow e} = \frac{2}{9}\,,\quad p_{e \rightarrow} = \frac{7}{9}\,,\quad p_{\leftarrow o} = \frac{1}{6}\,, \quad p_{o \rightarrow} = \frac{5}{6}\,.
\end{equation}
Now we need to find the probability that the operator at the end of the string, at the rightmost gate, is even or odd by relating it to the probability it was even or odd at the previous time step. We must take into account whether the operator has moved forward or backwards, as the probabilities of an even or odd operator appearing depend on which action occurs. When an even operator moves forward to an even operator, the rightmost operator will be either $X\iden $ or $Z \iden$, which occurs 6/7 times, the 1/7 occurs when an even operator moves forward to an odd operator, which only happens when $YY$ is generated and the operator at the rightmost gate is $Y\iden$. An even operator moves back to an even operator with probability $3/4$, when $X\iden$ or $Z\iden$ is generated and the operator coming from inside the circuit is $\{\iden, X, Z\}$. Here we have assumed that the 1-site operators coming from behind the evolving end of the string appear with uniform probability as the interior of the evolving Pauli string rapidly mixes. A similar analysis holds for the odd operators. All together we find the conditional probabilities 
\begin{align}
&p_{e\rightarrow e} = \frac{6}{7}\,,\quad p_{e\rightarrow o} = \frac{1}{7}\,, \quad p_{o \rightarrow e} = \frac{2}{5}\,,\quad p_{o\rightarrow o} = \frac{3}{5}\,,\nn
&p_{e\leftarrow e} = \frac{3}{4}\,,\quad p_{o\leftarrow e} = \frac{1}{4}\,, \quad p_{e \leftarrow o} = \frac{1}{4}\,,\quad p_{o\leftarrow o} = \frac{3}{4}\,.
\end{align}
The probability of the rightmost operator being even or odd at time $t$ is thus
\begin{align}
p_e(t) &= (p_{e\rightarrow e}p_{e\ra} + p_{e\leftarrow e}p_{e\la}) p_e(t-1) + (p_{o\rightarrow e}p_{o\ra} + p_{o\leftarrow e}p_{o\la}) p_o(t-1)\nn
p_o(t) &= (p_{o\rightarrow o}p_{o\ra} + p_{o\leftarrow o}p_{o\la}) p_o(t-1) + (p_{e\rightarrow o}p_{e\ra} + p_{e\leftarrow o}p_{e\la}) p_e(t-1)\,,
\label{eq:EOevol}
\end{align}
and for qubits we get the equations governing the evolving probabilities:
\begin{equation}
p_e(t) = \frac{5}{6} p_e(t-1) + \frac{3}{8} p_o(t-1)\,, \qquad p_o(t) = \frac{5}{8} p_o(t-1) + \frac{1}{6} p_e(t-1)\,.
\end{equation}
This itself is a Markov process, where the probability of finding an even or an odd operator at the end of the evolving string updates at each time step. We solve for the stationary distribution (\ie where the probabilities are the same at each time step) and find $p_e = 9/13$ and $p_o = 4/13$. Thus we can compute the probability of an operator moving back in terms of the probability the operator at the end of the evolving string is even or odd:
\begin{equation}
p = p_{\leftarrow e} p_e + p_{\leftarrow o} p_o\,,
\end{equation}
and find $p=8/39$ for qubits. This gives a butterfly velocity of $v_B = 23/39 \approx 0.5897$. Very close to the unitary butterfly velocity for qubits $v_B = 3/5$. In case the reader is suspect of how close the result is to the unitary case, that we have confirmed this value of the orthogonal butterfly velocity by computationally simulating the evolving operators in the random orthogonal circuits, shown in Fig.~\ref{fig:RQCvb}.

We can now derive the butterfly velocity for general local dimension. First we define the number of even or odd operators for a local dimension of $d$
\begin{equation}
n_e(d) = \frac{(d-1)(d+2)}{2}\,, \qquad n_o(d) = \frac{d(d-1)}{2}\,.
\end{equation}
Thus the number of even/odd operators at a single site is $n_e(q)$ and $n_o(q)$. The probabilities of an even and odd operator moving backwards are a simple generalization of what we discussed above in the case of qubits
\begin{equation}
p_{\leftarrow e} = \frac{n_e(q)}{n_e(d)} \and  p_{\leftarrow o} = \frac{n_o(q)}{n_o(d)}\,.
\end{equation}
%p_{e \rightarrow} = \frac{n_e(d)-n_e(q)}{n_e(d)}\,,\quad \quad p_{o \rightarrow} = \frac{n_o(d)-n_o(q)}{n_o(d)}
The probabilities of even/odd operators moving forwards or backwards to even/odd operators are
\begin{align}
p_{e\rightarrow e} = \frac{(n_e(q)+1)n_e(q)}{n_e(d)}\,,~~ p_{e\rightarrow o} = \frac{n_o(q)^2}{n_e(d)}\,,~~ p_{o \rightarrow e} = \frac{(n_e(q)+1)n_o(q)}{n_o(d)}\,,~~ p_{o\rightarrow o} = \frac{n_e(q)n_o(q)}{n_o(d)}\,,\nn
p_{e\leftarrow e} = \frac{(n_e(q)+1)n_e(q)}{n_e(d) q^2}\,,~~ p_{e\leftarrow o} = \frac{n_o(q)^2}{n_o(d) q^2}\,,~~ p_{o \leftarrow e} = \frac{n_e(q)n_o(q)}{n_e(d)q^2}\,,~~ p_{o\leftarrow o} = \frac{(n_e(q)+1)n_o(q)}{n_o(d) q^2}\,.
\end{align}
The evolution equations for the states of the random walker given in Eq.~\eqref{eq:EOevol} for general local dimension $q$ become
\begin{align}
p_e(t) &= \frac{(q+2)(q^2+1)}{2q(q^2+2)} p_e(t-1) + \frac{q^2-1}{2q^2} p_o(t-1)\nn
p_o(t) &= \frac{q^2+1}{2q^2} p_o(t-1) + \frac{(q-1)(q^2-q+2)}{2q(q^2+2)} p_e(t-1)\,.
\end{align}
As a sanity check, the probabilities of $p_e(t-1)$ and $p_o(t-1)$ add to unity. Solving for the stationary distribution, we find $p_e$ and $p_o$ and then compute the probability of an operator moving back to be
\begin{equation}
p = \frac{q^2+q+2}{(q+1)(q^3+2q+1)}\,.
\end{equation}
This gives a butterfly velocity for orthogonal random circuits
\begin{equation}
v_B = \frac{q^4+q^3 + q-3}{(q+1)(q^3+2q+1)}\,,
\label{eq:vbO}
\end{equation}
which has a series expansion in $q^2$ as
\begin{equation}
v_B = 1- \frac{2}{q^2} + \frac{6}{q^5} +\ldots\,,
\end{equation}
the same as the unitary case up to second order in $1/q$ in Eq.~\eqref{eq:Uvbexp}. Note the absence of contribution at $1/q^4$. 

The correlated random walk analysis for the COE and CSE diffusion constants is not so easily generalized to the orthogonal case. We compute the diffusion constant in the orthogonal circuit by deriving the diffusion equation from the difference equations for the 4-state Markov chain on the increments of the $O(d)$ random walk. The derivation in App.~\ref{app:diff}, gives the orthogonal diffusion constant Eq.~\eqref{eq:4stateD}. For local qubits, we find $D\approx 0.31$, similar to the unitary diffusion constant. Operator growth in orthogonal random circuits proceeds very similarly to that in unitary random circuits, although the analysis is somewhat more tedious. 

\subsection{Operator growth in random symplectic circuits}\label{sec:Sprqc}
Consider now a circuit of 2-site Haar random symplectic gates. To understand the operator growth, first consider the action of a single random gate
\begin{equation}
\int_{Sp(d)} dU\, \Tr\big(U \op_a U^\dagger \op_p\big) \Tr\big(U \op_b U^\dagger \op_p\big)\,.
\end{equation}
Computing the second moment we find the transition matrix for the symplectic circuits
\begin{align}
S_{pa} = \prod_s \Big( \delta^{(s)}_{a,1} \delta^{(s)}_{p,1} + \frac{1}{(d+1)(d-2)} \big(  &(\delta^{(s)}_{a,1}-1)(\delta^{(s)}_{p,1}-1) + (\delta^{(s)}_{a,1}-(-1)^{S_a})(\delta^{(s)}_{p,1}-(-1)^{S_p})\nn &- \frac{1}{d} (1-(-1)^{S_a})(1-(-1)^{S_p})\big) \Big)\,,
\end{align}
which is unsurprisingly similar to the orthogonal case. Noting that under symplectic conjugation Paulis are either even or odd: $\op_p^D \equiv J \op_p^T J^T = \pm \op_p$. Here we have defined $(-1)^{S_a}$ to be $+1$ for symplectically even Paulis and $-1$ for odd Paulis. The above transition matrix tells us that Paulis which are even under symplectic conjugation are taken to symplectically even Paulis and odd to odd.

In the symplectic case, we find that at each gate we take identities to identities, even non-identity Paulis to even and odd non-identity Paulis to odd:
\begin{align}
\begin{aligned}
\iden\,\iden &\ra \iden\,\iden &\qquad &\text{with prob}\quad 1 \nn
{\rm even}~ \op_p &\ra {\rm even}~\op_a & &\text{with prob}\quad p_e = 2/((d-1)(d+1)) \nn
{\rm odd}~ \op_p &\ra {\rm odd}~\op_a & &\text{with prob}\quad p_o = 2/(d(d+1))
\end{aligned}
\end{align}
where again we mean the even or odd action under symplectic conjugation $J \op_p^T J^T = \pm \op_p $. Note that this is not as simple as the even and odd operators in the orthogonal circuits. The matrix $J$ depends on the size of the gate, so for 2-qubit operators $J = iY \otimes \iden $. Even 2-qubit Paulis are $XY, \iden X, ZY \ldots$ and odd are $XZ, X\iden, \iden Y, \ldots $, which we see are not symmetric. 

\subsubsection*{Symplectic random walk}
The analysis for operator growth in symplectic random circuits is essentially the same as in the orthogonal case, we have that even and odd operators are mapped to themselves with uniform probability. The only difference is that the set of Paulis invariant under symplectic transpose is different that in the orthogonal case, thus giving slightly different probabilities. 
The evolution equations for the states of the symplectic random walker are still given by Eq.~\eqref{eq:EOevol}. Repeating the same analysis but with the symplectic version of $n_e$ and $n_o$, we find the probability of the end of an evolving Pauli string moving back to be
\begin{equation}
p= \frac{q^2-q+2}{(q^2+1) (q^2-q+1)}\,,
\end{equation}
which gives a butterfly velocity $v_B = 1-2p$
\begin{equation}
v_B = \frac{q^4-q^3+q-3}{(q^2+1) (q^2-q+1)} \quad \text{for qubits:} \quad v_B =\frac{7}{15}\,.
\label{eq:vbSp}
\end{equation}
In fact, the conditional probabilities of even/odd operators moving left/right, are different for the left/right edges of the growing operator, but both give the same butterfly velocity. The diffusion constant for the symplectic random walk is derived in App.~\ref{app:diff} from the 4-state Markov chain. From the probabilities for the symplectic random walk, we find $D\approx 0.31$.

\section{Discussion}
In this note, we studied operator growth in different classes of symmetric random circuits, with gates drawn from the orthogonal $O(d)$ and symplectic $Sp(d)$ groups, as well as the symmetric spaces $U(d)/O(d)$ (COE) and $U(d)/Sp(d)$ (CSE). These were models of local chaotic dynamics with antiunitary symmetries. Each class of random circuits gave rise to a different stochastic process on evolving Pauli strings, and operator growth could be solved by understanding the endpoint dynamics as a random walk. In the COE and CSE circuits, we understood operator growth as a persistent random walk with anticorrelation between steps. In the orthogonal and symplectic circuits, we found a similar random walk with bias and correlation, but where the random walker also carried an internal state. In all four of the classes of random circuits, we solved for the butterfly velocities and diffusion constants by solving the random walk, finding that symmetry slows down ballistic operator growth.

We can also comment more generally on properties of the random circuit models considered here. It is known that unitary circuits of polynomial depth form approximate $k$-designs \cite{HarrowLow08,Brandao12} and achieve optimal decoupling \cite{Brown15}. Haar-random $O(d)$ and $Sp(d)$ and random unitaries from $U(d)/O(d)$ and $U(d)/Sp(d)$ do not form $k$-designs \cite{NHJthesis,ChaosSym} (although the orthogonal and symplectic groups do form 1-designs). Thus symmetric circuits will not approximate moments of the unitary group. But one can define the notion of a symmetric $k$-design, reproducing moments of the invariant ensemble on the subgroup or subset of $U(d)$ \cite{NHJthesis,ChaosSym}. We expect that employing similar methods to \cite{HarrowLow08,Brandao12} would show that symmetric RQCs of polynomial depth form approximate symmetric $k$-designs. Moreover, random states with respect to the symmetric ensembles look locally maximally mixed and random symmetric unitaries achieve decoupling with only subleading corrections to the random unitary expression \cite{NHJthesis}. Thus, we expect that random symmetric circuits also achieve decoupling in polylogarithmic depth, but precisely computing the depth and understanding how suboptimal random symmetric circuits are would be an interesting avenue for future work.

Another interesting direction would be to consider symmetric random circuits in higher dimensions. The unitary circuits exhibit Kardar-Parisi-Zhang (KPZ) universality in the scaling in their front dynamics \cite{Nahum17}, with a diffusive broadening of $t^{1/3}$ in 2D circuits. As operator growth in 1D is somewhat constrained, it would interesting to see if the KPZ behavior persists in higher dimensional random symmetric circuits.

\subsection*{Acknowledgments}
The author thanks Yoni BenTov, Thom Bohdanowicz, Fernando Brand\~ao, Adam Nahum, John Preskill, Alex Turzillo, Amy Zhang, and especially Elizabeth Crosson and Beni Yoshida, for enlightening discussions and comments. NHJ acknowledges support from the DOE under award number {\rm DE-SC0018407}, as well as the Institute for Quantum Information and Matter (IQIM), an NSF Physics Frontiers Center (NSF Grant PHY-1733907), and the Perimeter Institute. Research at Perimeter Institute is supported by the Government of Canada through the Department of Innovation, Science and Economic Development Canada and by the Province of Ontario through the Ministry of Research, Innovation and Science.

\appendix

\section{Biased random walks and diffusion}\label{app:RW}
We can understand the derivation of the butterfly velocity and diffusion constant in the unitary random circuit as a simple biased random walk. Let's consider the right endpoint of the growing operator, so the bias refers to the preference of the operator to move right. At each time step we have a random variable $x_i$ which takes values $\pm 1$: we get $- 1$ with probability $p$, meaning the random walker moves left, and $+1$ with probability $1-p$, meaning the operator moves right. The position of the operator at a time $t$ is then given by
\begin{equation}
X(t) = \sum_{i=1}^t x_i\,.
\end{equation}
As the endpoint dynamics are Markovian, meaning the steps at successive times are uncorrelated and the random variables $x_i$ are iid, we simply compute the mean
\begin{equation}
\vev{X(t)} = \sum_{i=1}^t \vev{x_i} = t\vev{x_i} = t(1-2p) \quad\ra\quad v_B = 1-2p\,,
\end{equation}
which gives us the butterfly velocity $v_B$, and the second moment
\begin{equation}
\vev{X(t)^2} = \sum_{i,j=1}^t \vev{x_i x_j} = t\vev{x_i^2} + (t^2-t)\vev{x_i}\vev{x_j} = t + (t^2-t)(1-2p)^2\,,
\end{equation}
which gives the variance and the diffusion constant 
\begin{equation}
\vev{X(t)^2}_c = 4pt(1-p) = 2Dt \with D = 2p(1-p)\,.
\end{equation}
Therefore, computing the probability of moving back $p=1/(q^2+1)$ in the random circuits gives the above velocity and diffusion constants. 

We can arrive at the same conclusion by counting the left and right moves. Again, consider the motion of the right end of the Pauli string, and let $\ell$ denote the number of left steps and $r$ the number of right moves. Clearly, $t=\ell+r$ and the position of the random walker is $x=r-\ell$. The distribution on the number of left and right moves is
\begin{equation}
f(\ell,r) = \binom{t}{r} (1-p)^r p^\ell\,.
\end{equation}
From this we can compute the mean and variance of $x = r-\ell$ after $t$ timesteps get $v_B$ and $D$ as above. Said equivalently, the binomial distribution on the endpoint of the string is simply the sum of Bernoulli distributed random variables making up the steps in the random walk. At long times this gives rise to Eq.~\eqref{eq:bde}.

We will present a different derivation of the biased diffusion equation for the random walk as a warm-up for the derivations in App.~\ref{app:diff}, largely following \cite{WeissRW,Gardiner04}. The probability that the random walker is at site $x$ after $t$ steps obeys the evolution equation
\begin{equation}
\rho(x,t+1) = (1-p) \rho(x-1,t) + p \rho(x-1,t)\,.
\end{equation}
To take a continuum limit, we introduce spatial and temporal step sizes $\delta x$ and $\delta t$, then Taylor expand around small $\delta x$, $\delta t$, and find
\begin{equation}
\rho + \delta t \dot \rho + \ldots = (1-p) \Big(\rho - \delta x \rho' +\frac{\delta x^2}{2} \rho'' + \ldots\Big) + p \Big(\rho + \delta x \rho' +\frac{\delta x^2}{2} \rho'' + \ldots\Big) \,.
\end{equation}
Keeping only the lowest order terms in $\delta t$ and to second order in $\delta x$, we find
\begin{equation}
\dot \rho = -\frac{\delta x}{\delta t}(1-2p) \rho' +\frac{\delta x^2}{2\delta t} \rho''\,.
\end{equation}
Taking the limit $\delta x, \delta t \ra 0$, we identify the prefactor of the $\rho'$ term as velocity, $v_B = \lim_{\delta x, \delta t\ra 0} (1-2p) \delta x/\delta t$, and the prefactor of the $\rho''$ term as the diffusion constant, $D = \lim_{\delta x, \delta t\ra 0} \delta x^2/2\delta t$. Note that unlike in the case of the unbiased random walk, which simply yields the diffusion equation, here we need to be careful in how we take the limit such that both terms are finite in the continuum limit.  To ensure a sensible limit, we must require that $2D (1-2p) = v \delta x$ \cite{WeissRW}.
%Said equivalently, the bias must be close to 1/2 for the diffusion limit to hold. 
The result is the biased diffusion equation
\begin{equation}
\p_t \rho(x,t) = - v_B \p_x \rho(x,t) + D \p_x^2 \rho(x,t) \,, \where v_B = 1-2p ~~{\rm and}~~ D=2p(1-p)\,.
\end{equation}

\section{Difference equation derivations of diffusion} \label{app:diff}
In this appendix we briefly present a difference equation approach to computing $v_B$ and $D_\rho$ in the four classes of random circuits. For the COE and CSE random circuits, this serves as a check of the derivation in Sec.~\ref{sec:COErqc} and \ref{sec:CSErqc}, reproducing the same constants. But for the $O(d)$ and $Sp(d)$ random circuits, this is how we derive the diffusion constants. 

In the COE and CSE random circuits, we have a two-state Markov chain governing the edge evolution as the random walker has two different probabilities of moving back, depending on whether they moved back at the previous step. This is the persistence or inertia of the random walker. The random walker is thus in one of two states, forward state $f$ or back state $b$ depending on the previous step taken. $f(x,t)$ and $b(x,t)$ are the probabilities that the random walker moved forward/backward to site $x$ after $t$ steps. The two-state Markov chain $\{ f, b\}$ has the transition matrix
\begin{equation}
P = \begin{pmatrix} 1-p_1& p_1\\ 1-p_2& p_2\end{pmatrix}\,,
\end{equation}
where $p_1$ and $p_2$ are, respectively, the probabilities of moving back when in the forward and back states, as defined in Sec.~\ref{sec:COErqc}. We can write the two-state evolution as the difference equations
\begin{align}
f(x,t+1) &= (1-p_1) f(x-1,t) + (1-p_2) b(x-1,t)\nn
b(x,t+1) &= p_1 f(x+1,t) + p_2 b(x+1,t)\,,
\end{align}
introducing step sizes $\delta x$ and $\delta t$, we can Taylor expand and find
\begin{align}
f + \delta t \dot f + \ldots &= (1-p_1) \Big(f-\delta xf'+\frac{\delta x^2}{2} f'' +\ldots\Big) + (1-p_2) \Big(b-\delta xb'+\frac{\delta x^2}{2} b'' +\ldots\Big)\nn
b + \delta t \dot b + \ldots &= p_1 \Big(f+\delta xf'+\frac{\delta x^2}{2} f'' +\ldots\Big) + p_2 \Big(b+\delta xb'+\frac{\delta x^2}{2} b'' +\ldots\Big) \,.
\end{align}
Taking the sum and difference of the two equations, we can solve for the equation governing the evolution of $\rho = f+b$
\begin{equation}
\dot \rho = -\frac{\delta x}{\delta t} \bigg(\frac{1-p_1-p_2}{1-p_2+p_1}\bigg) \rho' + \frac{\delta x^2}{2\delta t} \bigg(\frac{1+p_2-p_1}{1-p_2+p_1}\bigg) \rho''\,,
\end{equation}
then taking the continuum limit $\delta x, \delta t \ra 0$ as described in App.~\ref{app:RW}, identifying $\delta x^2/2\delta t$ as $D_0$, we recover the biased diffusion equation $\p_t \rho = -v_B \p_x \rho + D \p_x^2 \rho$ with
\begin{equation}
v_B = 1-2p \and D = \frac{1+\alpha}{1-\alpha} D_0\,,
\end{equation}
and with persistence $\alpha = p_2-p_1$, uncorrelated diffusion constant $D_0 = 2p(1-p)$, and $p = p_1/(1-\alpha)$. This reproduces the result derived in Sec.~\ref{sec:COErqc}.

In the orthogonal and symplectic random circuits, the probabilities of moving forwards and backwards depend on whether the operator at the edge is even or odd. Thus the random walker has an internal state which is updated at every time step, being either in an even or odd state, as well as a forwards and backwards state. We have a four state Markov process governing the evolution of the random walker, with states $\{f_e, f_o, b_e, b_o\}$ and the transition matrix
\begin{equation}
P = \begin{pmatrix} \alpha& \beta& \gamma& \delta\\ \alpha& \beta& \gamma& \delta\\ \mu& \nu& \sigma& \tau\\ \mu& \nu& \sigma& \tau \end{pmatrix}\,, \where \begin{array}{c} \alpha+\beta+\gamma+\delta = 1\\ \mu+\nu+\sigma+\tau=1\end{array}\,.
\end{equation}
Here $\alpha$ is the probability of an even operator moving forward to an even operator, $\beta$ moving back to an even operator, etc. In the notation in Sec.~\ref{sec:Orqc}
\begin{align}
&\alpha = p_{e\ra} p_{e\ra e}\,, \quad \beta = p_{\la e} p_{e\la e}\,, \quad \gamma = p_{e\ra} p_{e\ra o}\,, \quad \delta =  p_{\la e} p_{o \la e}\nn
&\mu = p_{o\ra} p_{o\ra o}\,, \quad \nu = p_{\la o} p_{o\la o}\,, \quad \sigma = p_{o\ra} p_{o\ra e}\,, \quad \tau =  p_{\la o} p_{e \la o}\,.
\label{eq:cprobs}
\end{align}
We can again write down the difference equations
\begin{align}
f_e(x,t+1) &= \alpha f_e(x-1,t) + \alpha b_e(x-1,t) + \sigma f_o(x-1,t) + \sigma b_o(x-1,t) \nn
f_o(x,t+1) &= \gamma f_e(x-1,t) + \gamma b_e(x-1,t) + \mu f_o(x-1,t) + \mu b_o(x-1,t) \nn
b_e(x,t+1) &= \beta f_e(x+1,t) + \beta b_e(x+1,t) + \tau f_o(x+1,t) + \tau b_o(x+1,t) \nn
b_o(x,t+1) &= \delta f_e(x+1,t) + \delta b_e(x+1,t) + \nu f_o(x+1,t) + \nu b_o(x+1,t)\,,
\end{align}
and Taylor expand around $\delta x$ and $\delta t$, keeping only the terms to the first few orders  
{\small \begin{align}
\hspace*{-14pt} f_e + \delta t \dot f_e &= \alpha \Big(f_e-\delta xf_e'+\frac{\delta x^2}{2} f_e'' \Big) +\alpha \Big(b_e-\delta xb_e'+\frac{\delta x^2}{2} b_e'' \Big) + \sigma \Big(f_o-\delta xf_o'+\frac{\delta x^2}{2} f_o'' \Big) + \sigma\Big(b_o-\delta xb_o'+\frac{\delta x^2}{2} b_o'' \Big)\nn
\hspace*{-14pt} f_o + \delta t \dot f_o &= \gamma \Big(f_e-\delta xf_e'+\frac{\delta x^2}{2} f_e'' \Big) +\gamma \Big(b_e-\delta xb_e'+\frac{\delta x^2}{2} b_e'' \Big) + \mu \Big(f_o-\delta xf_o'+\frac{\delta x^2}{2} f_o'' \Big) + \mu\Big(b_o-\delta xb_o'+\frac{\delta x^2}{2} b_o'' \Big)\nn
\hspace*{-14pt} b_e + \delta t \dot b_e &= \beta \Big(f_e+\delta xf_e'+\frac{\delta x^2}{2} f_e'' \Big) +\beta \Big(b_e+\delta xb_e'+\frac{\delta x^2}{2} b_e'' \Big) + \tau \Big(f_o+\delta xf_o'+\frac{\delta x^2}{2} f_o'' \Big) + \tau\Big(b_o+\delta xb_o'+\frac{\delta x^2}{2} b_o'' \Big)\nn
\hspace*{-14pt} b_o + \delta t \dot b_o &= \delta \Big(f_e+\delta x f_e'+\frac{\delta x^2}{2} f_e'' \Big) +\delta \Big(b_e+\delta x b_e'+\frac{\delta x^2}{2} b_e'' \Big) + \nu \Big(f_o+\delta x f_o'+\frac{\delta x^2}{2} f_o'' \Big) + \nu\Big(b_o+\delta x b_o'+\frac{\delta x^2}{2} b_o'' \Big) \nonumber
\end{align}}
From the difference equations above, we can solve for the differential equation governing the evolution of $\rho(x,t) = f_e + f_o + b_e + b_o$, the probability of being at site $x$ after $t$ steps. The sum of the difference equations gives
\begin{equation}
\delta t \dot \rho = \delta x \big( (\beta+\delta-\alpha-\gamma) (f_e+b_e) + (\tau+\nu-\sigma -\mu) (f_o+b_o)\big) +\frac{\delta x^2}{2} \rho''\,.
\end{equation}
%Further, taking the difference of the above equations, we find
%\begin{equation}
%t \dot \rho = x \big( (\beta+\delta-\alpha-\gamma) (f_e+b_e) + (\tau+\nu-\sigma -\mu) (f_o+b_o)\big) +\frac{x^2}{2} \rho''\,.
%\end{equation}
Taking different linear combinations of the difference equations, we solve for the evolution of $\rho$. After extensive algebra, we recover the biased diffusion equation in the continuum limit for the 4-state random walk
%\begin{align}
%\dot \rho &= -\frac{\delta x}{\delta t} \bigg( \frac{(\gamma+\delta)(\sigma+\mu-\tau-\nu) + (\sigma+\tau)(\alpha+\gamma-\beta-\delta)}{\gamma+\delta+\sigma+\tau}\bigg) \rho'\\
% &\quad + \frac{\delta x^2}{2\delta t} \bigg( 1+ \frac{2(\alpha+\gamma-\mu-\sigma)}{\gamma+\delta+\sigma+\tau} \bigg( \frac{(\sigma+\tau)(\alpha-\gamma-\beta+\delta)+(\delta+\gamma)(\sigma-\mu-\tau+\nu))}{\gamma+\delta+\sigma+\tau}\bigg)\bigg) \nonumber
%\end{align}
with the butterfly velocity
\begin{equation}
v_B = \frac{(\gamma+\delta)(\sigma+\mu-\tau-\nu) + (\sigma+\tau)(\alpha+\gamma-\beta-\delta)}{\gamma+\delta+\sigma+\tau}
\label{eq:4statevb}
\end{equation}
and diffusion constant
\begin{equation}
D = \bigg( 1+ \frac{2(\alpha+\gamma-\mu-\sigma)}{\gamma+\delta+\sigma+\tau} \frac{(\sigma+\tau)(\alpha-\gamma-\beta+\delta)+(\delta+\gamma)(\sigma-\mu-\tau+\nu))}{\gamma+\delta+\sigma+\tau}\bigg)  D_0\,,
\label{eq:4stateD}
\end{equation}
where $D_0 = 2p(1-p)$. As a check, we plug the probabilities in Eq.~\eqref{eq:cprobs} for the $O(d)$ and $Sp(d)$ circuits into Eq.~\eqref{eq:4statevb} and recover the butterfly velocities derived in Sec.~\ref{sec:Orqc} and \ref{sec:Sprqc}. Plugging the probabilities into Eq.~\eqref{eq:4stateD} gives the diffusion constants for orthogonal and symplectic circuits.

\section{Haar integrals}\label{app:Haar}
We quickly review integration over the unitary group as well as other compact Lie groups and symmetric spaces. General moments of Haar-random unitaries can be written as \cite{Collins02,Collins04}
\begin{equation}
\int_{U(d)} dU\, U_{i_1 j_1}\ldots U_{i_k j_k} U^\dagger_{\ell_1 m_1}\ldots U^\dagger_{\ell_k m_k} = \sum_{\sigma,\tau\in S_k} \delta_\sigma (\vec \imath\, | \vec m) \delta_\tau (\vec \jmath\, | \vec \ell\, ) \Wg^U (\sigma^{-1} \tau,d)\,,
\end{equation}
where we sum over elements of $S_k$ and denote a $\delta$-function contraction of indices indexed by a permutation $\sigma \in S_k$ as $\delta_\sigma (\vec \imath\, | \vec \jmath\,) \equiv \delta_{i_1, j_{\sigma(1)}}\ldots\delta_{i_k, j_{\sigma(k)}}$. $\Wg^U(\sigma, d)$ is the unitary Weingarten function on a permutation $\sigma$, which can be computed from irreducible characters of $S_k$. The general expression for integration over Haar-random orthogonal matrices is \cite{Collins04,CollinsMat09}
\begin{equation}
\int dO\, O_{i_1 j_1} \ldots O_{i_{2k} j_{2k}} = \sum_{\sigma,\tau \in M_{2k}} \Delta_\sigma (\vec \imath\,) \Delta_\tau (\vec \jmath\,) \Wg^O (\sigma^{-1} \tau, d)\,,
\end{equation}
where we sum over the subset $M_{2k}$ of $S_{2k}$, corresponding to pair partitions, and define $\Delta_\sigma (\vec \imath\,) \equiv \delta_{i_{\sigma(1)}, i_{\sigma(2)}}\ldots\delta_{i_{\sigma(2k-1)}, i_{\sigma(2k)}}$ with $\sigma \in S_{2k}$, and $\Wg^O$ is the orthogonal Weingarten function. Similarly, integrals over Haar-random symplectic matrices can be written as \cite{Collins04,CollinsStolz08}
\begin{equation}
\int dS\, S_{i_1 j_1} \ldots S_{i_{2k} j_{2k}} = \sum_{\sigma,\tau \in M_{2k}} \Delta^J_\sigma (\vec \imath\,) \Delta^J_\tau (\vec \jmath\,) \Wg^{Sp} (\sigma^{-1} \tau, d)\,,
\end{equation}
where $\Delta^J_\sigma$ is a symplectic version of index contraction above with a $J$ inserted in each contraction and $\Wg^{Sp}$ is the symplectic Weingarten function. Note that $Sp(d)$ is the unitary symplectic group, the intersection of symplectic matrices with the unitary group, and thus $d$ must be taken to be even.
% (\ie with respect to the skew-symmetric bilinear form).

The circular orthogonal ensemble (COE) consists of the invariant measure on the compact symmetric space AI: $U(d)/O(d)$, realized as the subset of $U(d)$ consisting of symmetric unitaries. A random COE matrix is defined as $U^T U$ where $U$ is a Haar-random unitary. General expressions for integration over the COE were given in \cite{Matsumoto11,Matsumoto13} as
\begin{equation}
\int_{U(d)/O(d)} dU\, U_{i_1 i_2} \ldots U_{i_{2k-1} i_{2k}} U^\dagger_{j_1 j_2} \ldots U^\dagger_{j_{2k-1} j_{2k}} = \sum_{\sigma\in S_{2k}} \delta_\sigma (\vec \imath \,| \vec\jmath \,) \Wg^{\rm COE} (\sigma, d)\,,
\end{equation}
where the COE Weingarten functions are simply related to the orthogonal Weingarten functions. Similarly, integration over the circular symplectic ensemble (CSE), random instances of the compact symmetric space AII: $U(d)/Sp(d)$ corresponding to $U^D U$, is given by
\begin{equation}
\int_{U(d)/Sp(d)} dU\, J^T U_{i_1 i_2} \ldots  J^T U_{i_{2k-1} i_{2k}} J U^\dagger_{j_1 j_2} \ldots J U^\dagger_{j_{2k-1} j_{2k}} = \sum_{\sigma\in S_{2k}} \delta_\sigma (\vec \imath \,| \vec\jmath \,) \Wg^{\rm CSE} (\sigma, d)\,,
\end{equation}
where $\Wg^{\rm CSE}$ is simply related to the symplectic Weingarten functions. Again, here $d$ must be taken to be even.

\section{Random circuits for extended symmetry classes}
In this note, we discussed five classes of random circuits. As symmetry classes, these included the three Dyson classes for systems with time-reversal symmetry, the CUE, COE, and CSE ensembles, as well as the Haar measure on the orthogonal and symplectic groups. In the language of Altland and Zirnbauer \cite{AltlandZirnbauer}, fermionic systems go beyond the standard classification of symmetries and give a refinement into ten symmetry classes, depending on how the system realizes both time-reversal and particle-hole symmetry. The classes considered in this paper constitute five of those ten classes. The five additional ensembles include the three chiral ensembles and two BdG ensembles
\begin{equation*}
\begin{array}{l}
\text{AIII: } U(d)/(U(a)\times U(b))\\
\text{BDI: } O(d)/(O(a)\times O(b))\\
\text{CII: } Sp(d)/(Sp(a)\times Sp(b))
\end{array}
\begin{array}{l}
\text{DIII: } O(2d)/U(d)\\
\text{CI: } Sp(2d)/U(d)
\end{array}
\end{equation*}
with labels corresponding to Cartan's classification of compact symmetric spaces. Each of these quotients can be realized as the subset of $U(d)$ by the image of an involution on $U(d)$, from which an invariant probability measure is induced from the Haar measure. The Weingarten calculus to integrate over each of these spaces was developed in \cite{Matsumoto13}.

As extended symmetry classes arise in fermionic systems with time-reversal and particle-hole symmetry, we could consider toy models for these symmetry classes by constructing random circuits built from random unitaries drawn from one of these spaces. But unlike the Dyson symmetry classes, it is not necessarily true that generic interacting fermionic systems will fall into these classes, and in that sense are less universal. Nevertheless, we can compute the transition matrices for these additional five symmetry classes and find different `conservation' laws at each time step. For the first chiral ensemble, we find the action of the gate depends on the chiral transpose of the Pauli operator. But in general, the update rules we find are not so straightforward to interpret in all cases, and it is not so clear that each gives rise to a tractable random walk problem.

\section{Operator growth in random matrix theory}
Here we compute analytically compute the coefficients $\gamma_p(t)$ for the Gaussian unitary ensemble (GUE) in order to study operator growth in random matrix theory. Recall that we can expand any operator as $\op(t) = \sum_p \gamma_p(t) \op_p$, where the coefficients $\gamma_p(t)$ are the weights on a given Pauli string. 
%The fact that the operator norm is conserved means that the probabilities are conserved $\sum_p |\gamma_p(t)|^2 = 1$.
We want to consider evolving by a random matrix Hamiltonian, where $U = e^{-iHt}$ and $H \in $ GUE. To study operator growth we need to compute the coefficients $\gamma_p(t)$ averaged over GUE, more precisely we need the first and second moments of the coefficients. We will not review the necessary random matrix machinery here, but refer to \cite{ChaosRMT}. Consider the growth of an operator $\op_0(t) = e^{-iHt} \op_0 e^{iHt} $.
 
\subsubsection*{GUE averaged $\gamma_p(t)$}
We can compute the first moment
\begin{equation}
\gamma_p(t) = \frac{1}{d} \Tr \big( \op_0(t) \op_p \big) = \frac{1}{d} \Tr \big( e^{-iHt} \op_0 e^{iHt} \op_p \big)\,,
\end{equation}
using the unitary invariance of the GUE measure $dH$, then integrating using the 2nd Haar moment, and find
\begin{equation}
|\gamma_p(t)| = \frac{\CR_2 - 1}{d^2-1} \delta_{0,p} \quad\ra\quad |\gamma_{\op_0}(t)| \approx \frac{\CR_2(t)}{d^2}\,,
\end{equation}
with all other coefficients vanishing. Here $\CR_2(t) = \vev{|\Tr(e^{iHt})|^2}_H$ is the 2-point form factor for the GUE.
 
\subsubsection*{GUE averaged $|\gamma_p(t)|^2$}
More interesting is the average of $|\gamma_p(t)|^2$, which is the probability of finding a given operator $\op_p$ in the evolving operator. We can compute the second moment,
\begin{equation}
|\gamma_p(t)|^2 = \frac{1}{d^2} \Tr \big( e^{-iHt} \op_0 e^{iHt} \op_p \big)\Tr \big( e^{-iHt} \op_0 e^{iHt} \op_p \big)\,,
\end{equation}
using the unitary invariance of the GUE measure $dH$ and integrating using the 4th Haar moment. We find an expression involving $(4!)^2$ terms, which we will not reproduce here. First, we give the leading order behavior in $d$ which captures the early time piece, i.e. the decay of the support on the initial operator. For the late-time behavior we need the $1/d^2$ terms, as the coefficients all decay order $1/d^2$. At early-times, to leading order, we find
\begin{equation}
|\gamma_{\op_0}(t)|^2 \approx \frac{\CR_4(t)}{d^4} \and |\gamma_{p\neq \op_0}(t)|^2 \approx \frac{1}{d^2}\,.
\end{equation}
This isn't surprising; the support on the initial operator decays in time as $\CR_4(t) \sim 1/t^4$ and all other coefficients are around $1/d^2$ at early times. Here $\CR_4(t)$ is the 4-point form-factor.

Looking at the $1/d^2$ terms we can then discuss the late-time behavior of GUE operator growth. The coefficient of the initial operator $\op_0$ is
\begin{equation}
|\gamma_{\op_0}(t)|^2 \approx \frac{\CR_4}{d^4} +\frac{1}{d^2} \Big( 1- \frac{4\CR_2}{d^2} - \frac{4\CR_4}{d^4} - \frac{2\CR_{4,1}}{d^3} + \frac{\CR_{4,2}}{d^2} \Big)\,.
\end{equation}
Furthermore, we find that the probabilities of $\op_p$ depend on whether they commute or anticommute with the initial operator $\op_0$. The second moments of the coefficients are
\begin{align}
|\gamma_{p\neq \op_0}(t)|^2 \approx \frac{1}{d^2} - \frac{3\CR_4}{d^6} + \frac{2\CR_{4,1}}{d^5} \quad{\rm if}\quad [ \op_p, \op_0 ] = 0\,,\\
|\gamma_{p\neq \op_0}(t)|^2 \approx \frac{1}{d^2} + \frac{\CR_4}{d^6} - \frac{2\CR_{4,1}}{d^5} \quad{\rm if}\quad \{ \op_p, \op_0 \} = 0\,.
\end{align}
This means that at early times the $\op_0 $ coefficient decays from unity and all other coefficients are $\approx 1/d^2$. 
We note as a sanity check the non $\op_0$ coefficients vanish at $t=0$ as $\CR_4 = d^4$ and $\CR_{4,1} =d^3$. 
Around the dip time, when all the form factors are $\approx 1$, the coefficients are uniformly equal to $1/d^2$. At late-times there are $1/d^4$ fluctuations around the value $1/d^2$. The probabilities of the anticommuting operators increase with the ramp as the probabilities of the commuting operators decreases. The most interesting thing is that the initial operator becomes more likely again, with a probability twice that of the other operators. In summary,
\begin{equation}
{\rm Dip:}\quad |\gamma_{p}(t)|^2 \approx \frac{1}{d^2} \,, \qquad  {\rm Late:} \quad |\gamma_{\op_0}(t)|^2 \approx \frac{2}{d^2}  \and |\gamma_{p\neq \op_0}(t)|^2 \approx \frac{1}{d^2} \,.
\end{equation}
At the dip time, after the support on the initial operator has decayed, all operators are equally likely. But at late-times, in the plateau regime, the weight on the initial operator is twice that of all other operators.

\vspace*{8pt}
\bibliographystyle{utphys}
\bibliography{rqc_sym}

\providecommand{\href}[2]{#2}\begingroup\raggedright\begin{thebibliography}{10}

\bibitem{ELL05}
J.~Emerson, E.~Livine, and S.~Lloyd, ``Convergence conditions for random
  quantum circuits,'' \href{http://dx.doi.org/10.1103/PhysRevA.72.060302}{{\em
  Phys. Rev.} {\bfseries A72} (2005) 060302},
  \href{http://arxiv.org/abs/quant-ph/0503210}{{\ttfamily
  arXiv:quant-ph/0503210}}.

\bibitem{ODP07}
R.~Oliveira, O.~C.~O. Dahlsten, and M.~B. Plenio, ``Generic entanglement can be
  generated efficiently,''
  \href{http://dx.doi.org/10.1103/PhysRevLett.98.130502}{{\em Phys. Rev. Lett.}
  {\bfseries 98} (2007) 130502},
  \href{http://arxiv.org/abs/quant-ph/0605126}{{\ttfamily
  arXiv:quant-ph/0605126}}.

\bibitem{HarrowLow08}
A.~W. {Harrow} and R.~A. {Low}, ``{Random Quantum Circuits are Approximate
  2-designs},'' \href{http://dx.doi.org/10.1007/s00220-009-0873-6}{{\em Commun.
  Math. Phys.} {\bfseries 291} (2009) 257},
  \href{http://arxiv.org/abs/0802.1919}{{\ttfamily arXiv:0802.1919
  [quant-ph]}}.

\bibitem{Brandao12}
F.~G.~S.~L. {Brand{\~a}o}, A.~W. {Harrow}, and M.~{Horodecki}, ``{Local Random
  Quantum Circuits are Approximate Polynomial-Designs},''
  \href{http://dx.doi.org/10.1007/s00220-016-2706-8}{{\em Commun. Math. Phys.}
  {\bfseries 346} (2016) 397}, \href{http://arxiv.org/abs/1208.0692}{{\ttfamily
  arXiv:1208.0692 [quant-ph]}}.

\bibitem{Zni08}
M.~\v{Z}nidari\v{c}, ``Exact convergence times for generation of random
  bipartite entanglement,''
  \href{http://dx.doi.org/10.1103/PhysRevA.78.032324}{{\em Phys. Rev.}
  {\bfseries A78} (2008) 032324},
  \href{http://arxiv.org/abs/0809.0554}{{\ttfamily arXiv:0809.0554}}.

\bibitem{BV10}
W.~G. Brown and L.~Viola, ``Convergence rates for arbitrary statistical moments
  of random quantum circuits,''
  \href{http://dx.doi.org/10.1103/PhysRevLett.104.250501}{{\em Phys. Rev.
  Lett.} {\bfseries 104} (2010) 250501},
  \href{http://arxiv.org/abs/0910.0913}{{\ttfamily arXiv:0910.0913
  [quant-ph]}}.

\bibitem{Brown12}
W.~Brown and O.~Fawzi, ``{Scrambling speed of random quantum circuits},''
\href{http://arxiv.org/abs/1210.6644}{{\ttfamily arXiv:1210.6644 [quant-ph]}}.
%%CITATION = ARXIV:1210.6644;%%.

\bibitem{Brown15}
W.~{Brown} and O.~{Fawzi}, ``{Decoupling with random quantum circuits},''
  \href{http://dx.doi.org/10.1007/s00220-015-2470-1}{{\em Comm. Math. Phys.}
  {\bfseries 340} (2015) 867}, \href{http://arxiv.org/abs/1307.0632}{{\ttfamily
  arXiv:1307.0632 [quant-ph]}}.

\bibitem{Nahum17}
A.~Nahum, S.~Vijay, and J.~Haah, ``{Operator Spreading in Random Unitary
  Circuits},'' \href{http://dx.doi.org/10.1103/PhysRevX.8.021014}{{\em Phys.
  Rev.} {\bfseries X8} (2018) 021014},
\href{http://arxiv.org/abs/1705.08975}{{\ttfamily arXiv:1705.08975
  [cond-mat.str-el]}}.
%%CITATION = ARXIV:1705.08975;%%.

\bibitem{vonKey17}
C.~von Keyserlingk, T.~Rakovszky, F.~Pollmann, and S.~Sondhi, ``{Operator
  hydrodynamics, OTOCs, and entanglement growth in systems without conservation
  laws},'' \href{http://dx.doi.org/10.1103/PhysRevX.8.021013}{{\em Phys. Rev.}
  {\bfseries X8} (2018) 021013},
\href{http://arxiv.org/abs/1705.08910}{{\ttfamily arXiv:1705.08910
  [cond-mat.str-el]}}.
%%CITATION = ARXIV:1705.08910;%%.

\bibitem{Rakovszky17}
T.~Rakovszky, F.~Pollmann, and C.~W. von Keyserlingk, ``{Diffusive
  hydrodynamics of out-of-time-ordered correlators with charge conservation},''
\href{http://arxiv.org/abs/1710.09827}{{\ttfamily arXiv:1710.09827
  [cond-mat.stat-mech]}}.
%%CITATION = ARXIV:1710.09827;%%.

\bibitem{Khemani17}
V.~Khemani, A.~Vishwanath, and D.~A. Huse, ``{Operator spreading and the
  emergence of dissipation in unitary dynamics with conservation laws},''
\href{http://arxiv.org/abs/1710.09835}{{\ttfamily arXiv:1710.09835
  [cond-mat.stat-mech]}}.
%%CITATION = ARXIV:1710.09835;%%.

\bibitem{Nahum16}
A.~Nahum, J.~Ruhman, S.~Vijay, and J.~Haah, ``{Quantum Entanglement Growth
  Under Random Unitary Dynamics},''
  \href{http://dx.doi.org/10.1103/PhysRevX.7.031016}{{\em Phys. Rev.}
  {\bfseries X7} (2017) 031016},
\href{http://arxiv.org/abs/1608.06950}{{\ttfamily arXiv:1608.06950
  [cond-mat.stat-mech]}}.
%%CITATION = ARXIV:1608.06950;%%.

\bibitem{OpEnt18}
C.~Jonay, D.~A. Huse, and A.~Nahum, ``{Coarse-grained dynamics of operator and
  state entanglement},''
\href{http://arxiv.org/abs/1803.00089}{{\ttfamily arXiv:1803.00089
  [cond-mat.stat-mech]}}.
%%CITATION = ARXIV:1803.00089;%%.

\bibitem{RQCstatmech}
T.~Zhou and A.~Nahum, ``{Emergent statistical mechanics of entanglement in
  random unitary circuits},''
\href{http://arxiv.org/abs/1804.09737}{{\ttfamily arXiv:1804.09737
  [cond-mat.stat-mech]}}.
%%CITATION = ARXIV:1804.09737;%%.

\bibitem{LRbound}
E.~H. Lieb and D.~W. Robinson, ``{The finite group velocity of quantum spin
  systems},'' \href{http://dx.doi.org/10.1007/BF01645779}{{\em Commun. Math.
  Phys.} {\bfseries 28} (1972) 251}.

\bibitem{Hastings10}
M.~B. {Hastings},
  \href{http://dx.doi.org/10.1093/acprof:oso/9780199652495.001.0001}{``{Locality
  in Quantum Systems},''} in {\em {Quantum Theory from Small to Large Scales:
  Lecture Notes of the Les Houches Summer School}}, vol.~95.
\newblock 2010.
\newblock \href{http://arxiv.org/abs/1008.5137}{{\ttfamily arXiv:1008.5137
  [math-ph]}}.

\bibitem{HaydenPreskill}
P.~Hayden and J.~Preskill, ``{Black holes as mirrors: Quantum information in
  random subsystems},''
  \href{http://dx.doi.org/10.1088/1126-6708/2007/09/120}{{\em JHEP} {\bfseries
  09} (2007) 120},
\href{http://arxiv.org/abs/0708.4025}{{\ttfamily arXiv:0708.4025 [hep-th]}}.
%%CITATION = ARXIV:0708.4025;%%.

\bibitem{SekinoSusskind}
Y.~Sekino and L.~Susskind, ``{Fast Scramblers},''
  \href{http://dx.doi.org/10.1088/1126-6708/2008/10/065}{{\em JHEP} {\bfseries
  10} (2008) 065},
\href{http://arxiv.org/abs/0808.2096}{{\ttfamily arXiv:0808.2096 [hep-th]}}.
%%CITATION = ARXIV:0808.2096;%%.

\bibitem{FastScrambling}
N.~Lashkari, D.~Stanford, M.~Hastings, T.~Osborne, and P.~Hayden, ``{Towards
  the Fast Scrambling Conjecture},''
  \href{http://dx.doi.org/10.1007/JHEP04(2013)022}{{\em JHEP} {\bfseries 04}
  (2013) 022},
\href{http://arxiv.org/abs/1111.6580}{{\ttfamily arXiv:1111.6580 [hep-th]}}.
%%CITATION = ARXIV:1111.6580;%%.

\bibitem{LocalizedShocks}
D.~A. Roberts, D.~Stanford, and L.~Susskind, ``{Localized shocks},''
  \href{http://dx.doi.org/10.1007/JHEP03(2015)051}{{\em JHEP} {\bfseries 03}
  (2015) 051},
\href{http://arxiv.org/abs/1409.8180}{{\ttfamily arXiv:1409.8180 [hep-th]}}.
%%CITATION = ARXIV:1409.8180;%%.

\bibitem{RobertsLR}
D.~A. Roberts and B.~Swingle, ``{Lieb-Robinson Bound and the Butterfly Effect
  in Quantum Field Theories},''
  \href{http://dx.doi.org/10.1103/PhysRevLett.117.091602}{{\em Phys. Rev.
  Lett.} {\bfseries 117} (2016) 091602},
\href{http://arxiv.org/abs/1603.09298}{{\ttfamily arXiv:1603.09298 [hep-th]}}.
%%CITATION = ARXIV:1603.09298;%%.

\bibitem{Aleiner16}
I.~L. Aleiner, L.~Faoro, and L.~B. Ioffe, ``{Microscopic model of quantum
  butterfly effect: out-of-time-order correlators and traveling combustion
  waves},'' \href{http://dx.doi.org/10.1016/j.aop.2016.09.006}{{\em Annals
  Phys.} {\bfseries 375} (2016) 378},
\href{http://arxiv.org/abs/1609.01251}{{\ttfamily arXiv:1609.01251
  [cond-mat.stat-mech]}}.
%%CITATION = ARXIV:1609.01251;%%.

\bibitem{Gu16}
Y.~Gu, X.-L. Qi, and D.~Stanford, ``{Local criticality, diffusion and chaos in
  generalized Sachdev-Ye-Kitaev models},''
  \href{http://dx.doi.org/10.1007/JHEP05(2017)125}{{\em JHEP} {\bfseries 05}
  (2017) 125},
\href{http://arxiv.org/abs/1609.07832}{{\ttfamily arXiv:1609.07832 [hep-th]}}.
%%CITATION = ARXIV:1609.07832;%%.

\bibitem{PatelDiff}
A.~A. Patel, D.~Chowdhury, S.~Sachdev, and B.~Swingle, ``{Quantum butterfly
  effect in weakly interacting diffusive metals},''
  \href{http://dx.doi.org/10.1103/PhysRevX.7.031047}{{\em Phys. Rev.}
  {\bfseries X7} (2017) 031047},
\href{http://arxiv.org/abs/1703.07353}{{\ttfamily arXiv:1703.07353
  [cond-mat.str-el]}}.
%%CITATION = ARXIV:1703.07353;%%.

\bibitem{SwingleMPS}
S.~Xu and B.~Swingle, ``{Accessing scrambling using matrix product
  operators},''
\href{http://arxiv.org/abs/1802.00801}{{\ttfamily arXiv:1802.00801
  [quant-ph]}}.
%%CITATION = ARXIV:1802.00801;%%.

\bibitem{SYKopgrowth}
D.~A. Roberts, D.~Stanford, and A.~Streicher, ``{Operator growth in the SYK
  model},'' \href{http://dx.doi.org/10.1007/JHEP06(2018)122}{{\em JHEP}
  {\bfseries 06} (2018) 122},
\href{http://arxiv.org/abs/1802.02633}{{\ttfamily arXiv:1802.02633 [hep-th]}}.
%%CITATION = ARXIV:1802.02633;%%.

\bibitem{ScrambGraphs18}
G.~Bentsen, Y.~Gu, and A.~Lucas, ``{Fast scrambling on sparse graphs},''
\href{http://arxiv.org/abs/1805.08215}{{\ttfamily arXiv:1805.08215
  [cond-mat.str-el]}}.
%%CITATION = ARXIV:1805.08215;%%.

\bibitem{SYKopbeta}
X.-L. Qi and A.~Streicher, ``{Quantum Epidemiology: Operator Growth, Thermal
  Effects, and SYK},''
\href{http://arxiv.org/abs/1810.11958}{{\ttfamily arXiv:1810.11958 [hep-th]}}.
%%CITATION = ARXIV:1810.11958;%%.

\bibitem{DysonSym}
F.~J. {Dyson}, ``{The Threefold Way. Algebraic Structure of Symmetry Groups and
  Ensembles in Quantum Mechanics},''
  \href{http://dx.doi.org/10.1063/1.1703863}{{\em J. Math. Phys.} {\bfseries 3}
  (1962) 1199}.

\bibitem{LowThesis}
R.~A. {Low}, ``{Pseudo-randomness and Learning in Quantum Computation},''
  \href{http://arxiv.org/abs/1006.5227}{{\ttfamily arXiv:1006.5227
  [quant-ph]}}. PhD Thesis, 2010.

\bibitem{MSSbound}
J.~Maldacena, S.~H. Shenker, and D.~Stanford, ``{A bound on chaos},''
  \href{http://dx.doi.org/10.1007/JHEP08(2016)106}{{\em JHEP} {\bfseries 08}
  (2016) 106},
\href{http://arxiv.org/abs/1503.01409}{{\ttfamily arXiv:1503.01409 [hep-th]}}.
%%CITATION = ARXIV:1503.01409;%%.

\bibitem{SSbutterfly}
S.~H. Shenker and D.~Stanford, ``{Black holes and the butterfly effect},''
  \href{http://dx.doi.org/10.1007/JHEP03(2014)067}{{\em JHEP} {\bfseries 03}
  (2014) 067},
\href{http://arxiv.org/abs/1306.0622}{{\ttfamily arXiv:1306.0622 [hep-th]}}.
%%CITATION = ARXIV:1306.0622;%%.

\bibitem{SSstringy}
S.~H. Shenker and D.~Stanford, ``{Stringy effects in scrambling},''
  \href{http://dx.doi.org/10.1007/JHEP05(2015)132}{{\em JHEP} {\bfseries 05}
  (2015) 132},
\href{http://arxiv.org/abs/1412.6087}{{\ttfamily arXiv:1412.6087 [hep-th]}}.
%%CITATION = ARXIV:1412.6087;%%.

\bibitem{ChaosChannels}
P.~Hosur, X.-L. Qi, D.~A. Roberts, and B.~Yoshida, ``{Chaos in quantum
  channels},'' \href{http://dx.doi.org/10.1007/JHEP02(2016)004}{{\em JHEP}
  {\bfseries 02} (2016) 004},
\href{http://arxiv.org/abs/1511.04021}{{\ttfamily arXiv:1511.04021 [hep-th]}}.
%%CITATION = ARXIV:1511.04021;%%.

\bibitem{ChaosDesign}
D.~A. Roberts and B.~Yoshida, ``{Chaos and complexity by design},''
  \href{http://dx.doi.org/10.1007/JHEP04(2017)121}{{\em JHEP} {\bfseries 04}
  (2017) 121},
\href{http://arxiv.org/abs/1610.04903}{{\ttfamily arXiv:1610.04903
  [quant-ph]}}.
%%CITATION = ARXIV:1610.04903;%%.

\bibitem{ChaosRMT}
J.~Cotler, N.~Hunter-Jones, J.~Liu, and B.~Yoshida, ``{Chaos, Complexity, and
  Random Matrices},'' \href{http://dx.doi.org/10.1007/JHEP11(2017)048}{{\em
  JHEP} {\bfseries 11} (2017) 048},
\href{http://arxiv.org/abs/1706.05400}{{\ttfamily arXiv:1706.05400 [hep-th]}}.
%%CITATION = ARXIV:1706.05400;%%.

\bibitem{KhemaniVel18}
V.~Khemani, D.~A. Huse, and A.~Nahum, ``{Velocity-dependent Lyapunov exponents
  in many-body quantum, semiclassical, and classical chaos},''
  \href{http://dx.doi.org/10.1103/PhysRevB.98.144304}{{\em Phys. Rev.}
  {\bfseries B98} (2018) 144304},
\href{http://arxiv.org/abs/1803.05902}{{\ttfamily arXiv:1803.05902
  [cond-mat.stat-mech]}}.
%%CITATION = ARXIV:1803.05902;%%.

\bibitem{ChaosSym}
N.~Hunter-Jones and B.~Yoshida, ``{Late-time chaos and symmetry}.'' {\it In
  preparation}.

\bibitem{NHJthesis}
N.~Hunter-Jones, \href{http://dx.doi.org/10.7907/BHZ5-HV76}{{\em {Chaos and
  Randomness in Strongly-Interacting Quantum Systems}}}.
\newblock PhD thesis, California Institute of Technology, 2018.

\bibitem{Collins02}
B.~Collins, ``{Moments and cumulants of polynomial random variables on unitary
  groups, the Itzykson-Zuber integral, and free probability},''
  \href{http://dx.doi.org/10.1155/S107379280320917X}{{\em Int. Math. Res. Not.}
  {\bfseries 2003} (2003) 953},
  \href{http://arxiv.org/abs/math-ph/0205010}{{\ttfamily
  arXiv:math-ph/0205010}}.

\bibitem{Collins04}
B.~{Collins} and P.~{{\'S}niady}, ``{Integration with Respect to the Haar
  Measure on Unitary, Orthogonal and Symplectic Group},''
  \href{http://dx.doi.org/10.1007/s00220-006-1554-3}{{\em Commun. Math. Phys.}
  {\bfseries 264} (2006) 773},
  \href{http://arxiv.org/abs/math-ph/0402073}{{\ttfamily
  arXiv:math-ph/0402073}}.

\bibitem{CollinsMat09}
B.~Collins and S.~Matsumoto, ``On some properties of orthogonal weingarten
  functions,'' \href{http://dx.doi.org/10.1063/1.3251304}{{\em J. Math. Phys.}
  {\bfseries 50} (2009) 113516},
  \href{http://arxiv.org/abs/0903.5143}{{\ttfamily arXiv:0903.5143 [math-ph]}}.

\bibitem{Matsumoto11}
S.~{Matsumoto}, ``{General moments of matrix elements from circular orthogonal
  ensembles},'' \href{http://dx.doi.org/10.1142/S2010326312500050}{{\em Random
  Matrices: Theory Appl.} {\bfseries 01} (2012) 1250005},
  \href{http://arxiv.org/abs/1109.2409}{{\ttfamily arXiv:1109.2409 [math.PR]}}.

\bibitem{Matsumoto13}
S.~{Matsumoto}, ``{Weingarten calculus for matrix ensembles associated with
  compact symmetric spaces},''
  \href{http://dx.doi.org/10.1142/S2010326313500019}{{\em Random Matrices:
  Theory Appl.} {\bfseries 02} (2013) 1350001},
  \href{http://arxiv.org/abs/1301.5401}{{\ttfamily arXiv:1301.5401 [math.PR]}}.

\bibitem{Taylor22}
G.~I. Taylor, ``{Diffusion by Continuous Movements},''
  \href{http://dx.doi.org/10.1112/plms/s2-20.1.196}{{\em Proc. London Math.
  Soc.} {\bfseries 20} (1922) 196}.

\bibitem{GoldsteinDiff}
S.~Goldstein, ``On diffusion by discontinuous movements, and on the telegraph
  equation,'' \href{http://dx.doi.org/10.1093/qjmam/4.2.129}{{\em Quart. J.
  Mech. Appl. Math.} {\bfseries 4} (1951) 129}.

\bibitem{RenshawCorr}
E.~Renshaw and R.~Henderson, ``The correlated random walk,'' {\em J. Appl.
  Probab.} {\bfseries 18} (1981) 403.

\bibitem{WeissRW}
G.~H. Weiss, {\em Aspects and applications of the random walk}.
\newblock Random Materials and Processes Series. North-Holland, 1994.

\bibitem{Gardiner04}
C.~Gardiner, {\em Handbook of Stochastic Methods for Physics, Chemistry, and
  the Natural Sciences}.
\newblock Springer Complexity. Springer, 2004.

\bibitem{CollinsStolz08}
B.~Collins and M.~Stolz, ``Borel theorems for random matrices from the
  classical compact symmetric spaces,''
  \href{http://dx.doi.org/10.1214/07-AOP341}{{\em Ann. Probab.} {\bfseries 36}
  (2008) 876}, \href{http://arxiv.org/abs/math/0611708}{{\ttfamily
  arXiv:math/0611708 [math.PR]}}.

\bibitem{AltlandZirnbauer}
A.~Altland and M.~R. Zirnbauer, ``Nonstandard symmetry classes in mesoscopic
  normal-superconducting hybrid structures,''
  \href{http://dx.doi.org/10.1103/PhysRevB.55.1142}{{\em Phys. Rev.} {\bfseries
  B55} (1997) 1142}, \href{http://arxiv.org/abs/cond-mat/9602137}{{\ttfamily
  arXiv:cond-mat/9602137}}.

\end{thebibliography}\endgroup


\providecommand{\href}[2]{#2}\begingroup\raggedright\endgroup

\end{document}